\documentclass[twocolumn,showpacs,preprintnumbers,amsmath,amssymb,superscriptaddress]{revtex4}

\usepackage{graphicx}
\usepackage{dcolumn}
\usepackage{bm}
\usepackage[figuresright]{rotating}
\usepackage{fancyhdr}
\usepackage{enumerate}
\usepackage{indentfirst}
\usepackage{xspace}
\usepackage{color}
\usepackage{epsfig}
\usepackage{subfigure}

\newcommand{\etal}{{\it et\ al}.}

\newcommand{\nue}{\nu_e}
\newcommand{\numu}{\nu_\mu}
\newcommand{\nutau}{\nu_\tau}

\newcommand{\numunutau}{\nu_\mu \leftrightarrow \nu_\tau}

\newcommand{\eee}{\varepsilon_{ee}}
\newcommand{\eem}{\varepsilon_{e\mu}}
\newcommand{\eet}{\varepsilon_{e\tau}}
\newcommand{\emm}{\varepsilon_{\mu\mu}}
\newcommand{\emt}{\varepsilon_{\mu\tau}}
\newcommand{\ett}{\varepsilon_{\tau\tau}}

\begin{document}

\title{Study of Non-Standard Neutrino Interactions with Atmospheric Neutrino Data in Super-Kamiokande I and II}

\date{\today}

\newcommand{\AFFicrr}{\affiliation{Kamioka Observatory, Institute for Cosmic Ray Research, University of Tokyo, Kamioka, Gifu 506-1205, Japan}}
\newcommand{\AFFkashiwa}{\affiliation{Research Center for Cosmic Neutrinos, Institute for Cosmic Ray Research, University of Tokyo, Kashiwa, Chiba 277-8582, Japan}}
\newcommand{\AFFipmu}{\affiliation{Institute for the Physics and
Mathematics of the Universe, University of Tokyo, Kashiwa, Chiba
277-8582, Japan}}
\newcommand{\AFFbu}{\affiliation{Department of Physics, Boston University, Boston, MA 02215, USA}}
\newcommand{\AFFbnl}{\affiliation{Physics Department, Brookhaven National Laboratory, Upton, NY 11973, USA}}
\newcommand{\AFFucd}{\affiliation{Department of Physics, University of California, Davis, Davis, CA 95616, USA}}
\newcommand{\AFFuci}{\affiliation{Department of Physics and Astronomy, University of California, Irvine, Irvine, CA 92697-4575, USA }}
\newcommand{\AFFcsu}{\affiliation{Department of Physics, California State University, Dominguez Hills, Carson, CA 90747, USA}}
\newcommand{\AFFcnm}{\affiliation{Department of Physics, Chonnam National University, Kwangju 500-757, Korea}}
\newcommand{\AFFduke}{\affiliation{Department of Physics, Duke University, Durham NC 27708, USA}}
\newcommand{\AFFfukuoka}{\affiliation{Junior College, Fukuoka Institute of Technology, Fukuoka, Fukuoka 811-0295, Japan}}
\newcommand{\AFFgmu}{\affiliation{Department of Physics, George Mason University, Fairfax, VA 22030, USA }}
\newcommand{\AFFgifu}{\affiliation{Department of Physics, Gifu University, Gifu, Gifu 501-1193, Japan}}
\newcommand{\AFFuh}{\affiliation{Department of Physics and Astronomy, University of Hawaii, Honolulu, HI 96822, USA}}
\newcommand{\AFFkanagawa}{\affiliation{Physics Division, Department of Engineering, Kanagawa University, Kanagawa, Yokohama 221-8686, Japan}}
\newcommand{\AFFkek}{\affiliation{High Energy Accelerator Research Organization (KEK), Tsukuba, Ibaraki 305-0801, Japan }}
\newcommand{\AFFkobe}{\affiliation{Department of Physics, Kobe University, Kobe, Hyogo 657-8501, Japan}}
\newcommand{\AFFkyoto}{\affiliation{Department of Physics, Kyoto University, Kyoto, Kyoto 606-8502, Japan}}
\newcommand{\AFFumd}{\affiliation{Department of Physics, University of Maryland, College Park, MD 20742, USA }}
\newcommand{\AFFmit}{\affiliation{Department of Physics, Massachusetts Institute of Technology, Cambridge, MA 02139, USA}}
\newcommand{\AFFmiyagi}{\affiliation{Department of Physics, Miyagi University of Education, Sendai, Miyagi 980-0845, Japan}}
\newcommand{\AFFnagoya}{\affiliation{Solar Terrestrial Environment Laboratory, Nagoya University, Nagoya, Aichi 464-8602, Japan}}
\newcommand{\AFFsuny}{\affiliation{Department of Physics and Astronomy, State University of New York, Stony Brook, NY 11794-3800, USA}}
\newcommand{\AFFniigata}{\affiliation{Department of Physics, Niigata University, Niigata, Niigata 950-2181, Japan }}
\newcommand{\AFFokayama}{\affiliation{Department of Physics, Okayama University, Okayama, Okayama 700-8530, Japan }}
\newcommand{\AFFosaka}{\affiliation{Department of Physics, Osaka University, Toyonaka, Osaka 560-0043, Japan}}
\newcommand{\AFFseoul}{\affiliation{Department of Physics, Seoul National University, Seoul 151-742, Korea}}
\newcommand{\AFFshizuokasc}{\affiliation{Department of Informatics in
Social Welfare, Shizuoka University of Welfare, Yaizu, Shizuoka, 425-8611, Japan}}
\newcommand{\AFFshizuoka}{\affiliation{Department of Systems Engineering, Shizuoka University, Hamamatsu, Shizuoka 432-8561, Japan}}
\newcommand{\AFFskk}{\affiliation{Department of Physics, Sungkyunkwan University, Suwon 440-746, Korea}}
\newcommand{\AFFtohoku}{\affiliation{Research Center for Neutrino Science, Tohoku University, Sendai, Miyagi 980-8578, Japan}}
\newcommand{\AFFtokyo}{\affiliation{The University of Tokyo, Bunkyo, Tokyo 113-0033, Japan }}
\newcommand{\AFFtokai}{\affiliation{Department of Physics, Tokai University, Hiratsuka, Kanagawa 259-1292, Japan}}
\newcommand{\AFFtit}{\affiliation{Department of Physics, Tokyo Institute
for Technology, Meguro, Tokyo 152-8551, Japan }}
\newcommand{\AFFtsinghua}{\affiliation{Department of Engineering Physics, Tsinghua University, Beijing, 100084, China}}
\newcommand{\AFFwarsaw}{\affiliation{Institute of Experimental Physics, Warsaw University, 00-681 Warsaw, Poland }}
\newcommand{\AFFuw}{\affiliation{Department of Physics, University of Washington, Seattle, WA 98195-1560, USA}}

\AFFicrr
\AFFkashiwa
\AFFipmu
\AFFbu
\AFFbnl
\AFFuci
\AFFcsu
\AFFcnm
\AFFduke
\AFFfukuoka
\AFFgifu
\AFFuh
\AFFkanagawa
\AFFkek
\AFFkobe
\AFFkyoto
\AFFmiyagi
\AFFnagoya
\AFFsuny
\AFFniigata
\AFFokayama
\AFFosaka
\AFFseoul
\AFFshizuokasc
\AFFskk
\AFFtokai
\AFFtokyo
\AFFtsinghua
\AFFwarsaw
\AFFuw
%
\author{G.~Mitsuka}
\altaffiliation{Present address: Solar Terrestrial Environment Laboratory, Nagoya University, Nagoya, Aichi 464-8602, Japan}
\AFFkashiwa

\author{K.~Abe}
\AFFicrr
\author{Y.~Hayato}
\AFFicrr
\AFFipmu
\author{T.~Iida}
\author{M.~Ikeda}
\author{J.~Kameda}
\author{Y.~Koshio}
\author{M.~Miura} 
\AFFicrr
\author{S.~Moriyama} 
\author{M.~Nakahata} 
\AFFicrr
\AFFipmu
% ...

\collaboration{The Super-Kamiokande Collaboration}
\noaffiliation

%
%
\begin{abstract}
  In this paper we study non-standard neutrino interactions as an example of physics beyond the standard model 
using atmospheric neutrino data collected during the Super-Kamiokande I(1996-2001) and II(2003-2005) periods. 
We focus on flavor-changing-neutral-currents (FCNC), which allow neutrino flavor transitions via neutral current interactions, 
and effects which violate lepton non-universality (NU) and give rise to different neutral-current interaction-amplitudes for different neutrino flavors. 
We obtain a limit on the FCNC coupling parameter, $\emt$, $|\emt|<1.1\times10^{-2}$ at $90\%$C.L. 
and various constraints on other FCNC parameters as a function of the NU coupling, $\eee$. 
We find no evidence of non-standard neutrino interactions in the Super-Kamiokande atmospheric data. 
\end{abstract}


\maketitle

%
%
\section{Introduction}\label{chap:intro}
The experimental understanding of neutrino oscillations has improved dramatically over the last ten years. 
In 1998 Super-Kamiokande (``Super-K'') reported an up-down asymmetry in the zenith angle distribution of 
muon-like ($\mu$-like) events and concluded the distortion was evidence for neutrino oscillations~\cite{exp:SKevi}.
Super-K also observed an oscillation signature consistent with the $L/E$ (pathlength over energy) dependence
predicted by $\numu$ to $\nutau$ oscillations at maximal $\theta_{23}$ mixing~\cite{exp:skloe}. 
Results consistent with the $\numu \to \nutau$ oscillation hypothesis have also been obtained by the tau appearance analysis in Super-K~\cite{exp:sktau} and the long-baseline accelerator experiments K2K~\cite{exp:k2k} and MINOS~\cite{exp:minos}.

However, many alternatives to neutrino oscillations have been proposed to 
explain the asymmetry of the atmospheric neutrino $\mu$-like event sample~\cite{Lipari99}.
Neutrino decoherence~\cite{Grossman, Lisi}, neutrino decay~\cite{Barger1, Barger2}, 
mass-varying neutrinos~\cite{MaVan}
, and CPT violation effects are among the most prominent. 
Although most of these approaches have been ruled out by 
the Super-K~\cite{exp:skloe, exp:skmavan} and MINOS~\cite{exp:minos_cpt} data, 
non-standard neutrino interactions (NSI) with matter, that is interactions which are not predicted by the Standard Model,
remain viable.

Among the many types of NSI models, we focus in this paper on two generic types of neutrino interactions. 
Flavor-changing-neutral-current (FCNC) effects represent neutrino interactions with fermions, $f$, in matter that 
induce neutrino flavor change: $\nu_\alpha + f \rightarrow \nu_\beta + f$,  where $\alpha$ and $\beta$ denote neutrino flavors.
Lepton non-universal (NU) interactions, on the other hand, are defined by a non-universal neutral current 
scattering amplitude among the three flavored neutrinos. 
In the Standard Model, this amplitude is identical among the neutrinos. 
Other theorios of neutrino mass, however, predict these kinds of interactions. For instance,
NSI interactions are often seen in models where the neutrino mass arises from 
admixtures of isosinglet neutral heavy leptons~\cite{Valle80} or from $R$-parity violating supersymmetry~\cite{susy}.
A summary of various NSI models is presented in Ref.~\cite{Fornengo}. 
Although NSI are predicted by various theories, the expected phenomena do not generally depend on the particular phenomenological model
and are typically characterized by dependence on the neutrino energy and the surrounding matter density. 
For this reason, NSI can be explored in a general context using atmospheric neutrinos.
While the current constraints on NSI come from beam-based experiments~\cite{CHARM_NSI,NuTeV_NSI}, 
atmospheric neutrinos can provide additional sensitivity to these interactions due to their ample flux 
and the large amount of matter they traverse before detection. 

This paper discusses atmospheric neutrino oscillations in the context of NSI at Super-K and is organized as follows. 
The data set and oscillation framework used in this paper are presented in Sec.~\ref{chap:data}  and Sec.~\ref{chap:2dnsi_form}, respectively. 
In Sec.~\ref{chap:2d-hybrid} we show the results of an analysis 
assuming two-flavor $\numu \leftrightarrow \nutau$ neutrino oscillations amidst NSI. 
An analysis using an extended three-flavor framework is performed in Sec.~\ref{chap:3d-hybrid} and finally, 
the results are summarized in Sec.~\ref{chap:conclusions}.

%
%
\section{Atmospheric neutrino data}\label{chap:data}
Super-K is a 50 kiloton water Cherenkov detector located in a zinc mine in Kamioka, Japan. 
It is optically separated into an Inner Detector (ID) which is istrumented with 11,146 inward facing photomultiplier tubes (PMTs) 
at full capacity, and an outer detector (OD) used to veto cosmic ray muons.  
A more detailed description of the detector is presented in Ref.\cite{skfull}. 
The run period of Super-K has been classified into four phases: 
The first phase corresponds to physics data taken between April 1996 and July 2001 (SK-I). 
After an accident at the end of 2001, 
Super-K resumed data taking with half the number of ID PMTs between October 2002 and October 2005 (SK-II). 
The remaining two run periods are divided into SK-III (2006-2008), after rebuilding the ID with its full complement of PMTs, 
and SK-IV (2008-present), after the data acquisition system was upgraded. 
In this paper we use data from the SK-I and SK-II run periods.

\subsubsection{Classification of atmospheric neutrino data}\label{chap:classification}
Atmospheric neutrino events in Super-K are divided into the following four categories: 
fully contained\,(FC), partially contained\,(PC), upward stopping muons\,(UPMU stopping) and upward through-going muons\,(UPMU through-going). 
For FC and PC events, event vertices are required to be within the $\sim$22.5 kton fiducial volume defined by the volume 
inset from the ID walls by 2\,m.
An event whose particles are completely contained within the ID is classified as FC, while an event with particles 
exiting the ID and depositing energy into the outer detector (OD) is classified as PC. 
The PC sample is further classified into two sub-categories, PC stopping and PC through-going. 
The former corresponds to events with a particle that stops in the OD, 
while in the case of the latter, the particle exits the OD.
UPMU events are produced by the charged current interactions of atmospheric muon neutrinos 
in the rock surrounding the detector. 
Muons traveling in the upward direction are selected to avoid contamination from cosmic ray muons. 
The UPMU stopping sample is defined by events which enter from outside the detector and stop inside the ID, 
while the UPMU through-going sample is composed of those that enter and subsequently exit the ID. 
The expected mean energy for each of the event class is, 
$\sim$ 1\,GeV for FC, $\sim$ 10\,GeV for PC, $\sim$ 10\,GeV for UPMU stopping, and $\sim$ 100\,GeV for UPMU through-going. 
Descriptions of the event reduction and reconstruction can be found in~\cite{skfull}.

\subsubsection{Monte Carlo simulation}\label{chap:simulation}
In this paper, independent 500 year Monte Carlo (MC) samples are used for SK-I and SK-II. 
The analyses use the Honda2006 neutrino flux~\cite{flux:honda2006} and neutrino interactions are simulated using the NEUT interaction generator~\cite{neut_mitsuka}.

\section{Oscillation Framework}
\label{chap:2dnsi_form}

In the following sections we will consider two separate NSI models derived from a more general formalism.
We introduce first the more general framework, restricting its scope later to the particular NSI effects we aim to study. 
The full three-flavor Hamiltonian, $H_{\alpha\beta}$, governing the propagation of neutrinos in the presence 
of effects from NSI is 
\begin{widetext}
  \begin{equation}
    H_{\alpha\beta} = \frac{1}{2E} U_{\alpha j}
    \begin{pmatrix}
      0 & 0 & 0 \\
      0 & \Delta m^2_{21} & 0 \\
      0 & 0 & \Delta m^2_{31} \\
    \end{pmatrix}
    (U^\dagger)_{k \beta} +
    V_{\rm MSW}
    + \sqrt{2}G_FN_f
    \begin{pmatrix}
      \eee & \eem & \eet \\
      \eem & \emm & \emt \\
      \eet & \emt & \ett \\
    \end{pmatrix}.
    \label{eq:3dnsi_hamiltonian}
  \end{equation}
\end{widetext}
\noindent In this equation  $U$ is the unitary PMNS matrix~\cite{PMNS}, which describes standard neutrino mixing as rotations among 
pairs of mass eigenstates parameterized by unique mixing angles, $\theta_{ij}$.
Here the squared difference of the neutrino masses is denoted by  $\Delta m^2_{ij}$, 
$V_{\rm MSW}$ is the MSW potential in the flavor basis~\cite{MSW},  $G_F$ is the Fermi coupling constant, 
$N_f$ is the fermion number density in matter along the path of the neutrino, and the $\varepsilon_{\alpha\beta}$ represent 
the NSI coupling parameters. The non-universal couplings are represented by the flavor diagonal $\varepsilon$ and the FCNC interactions by the off-diagonal elements.
Standard neutrino oscillations are recovered when all of the $\varepsilon_{\alpha\beta} = 0$. 
Note that while the first term of the Hamiltonian carries an explicit energy dependence, the second and third terms do not and instead are functions of the local matter density. 
For our calculations below we employ the PREM model~\cite{PREM} of the Earth's density profile and chemical composition, 
where the proton to nucleon ratio in the mantle and core are set to be $Y_p$ = 0.497 and 0.468, respectively~\cite{Bahcall}. 
In many of our calculations we use the average matter density along the path of the neutrino.

%
%
\section{Analysis with a Two-Flavor Hybrid Model}\label{chap:2d-hybrid}
We consider first a model in which NSI effects in the $\numu-\nutau$
sector coexist with standard two-flavor $\numu \leftrightarrow \nutau$ neutrino oscillations.
In this scenario all NSI that couple to $\nu_e$ in Eq.(\ref{eq:3dnsi_hamiltonian}),
$\varepsilon_{e \beta}$, are set to zero. 
Allowing the remaining parameters which couple to $\numu$ and $\nutau$ to be non-zero introduces 
a matter-dependent effect on the oscillations of $\numu \leftrightarrow \nutau$. 
Since the standard two-flavor scenario ($\theta_{12}$, $\theta_{13}$, and $\Delta m^2_{21} = 0$) 
does not incorporate oscillations into $\nu_e$, there is no separate 
effect from the standard matter potential, $V_{\rm MSW}$.
Labeling this the \textit{2-flavor hybrid model}, we can explore NSI couplings by searching for matter-induced
distortions of standard  $\numu \leftrightarrow \nutau $ oscillations. 

\subsection{Formalism}

\begin{figure*}[htbp]
  \vspace{0.5cm}
  \begin{center}
    \includegraphics[width=7.0cm,keepaspectratio]{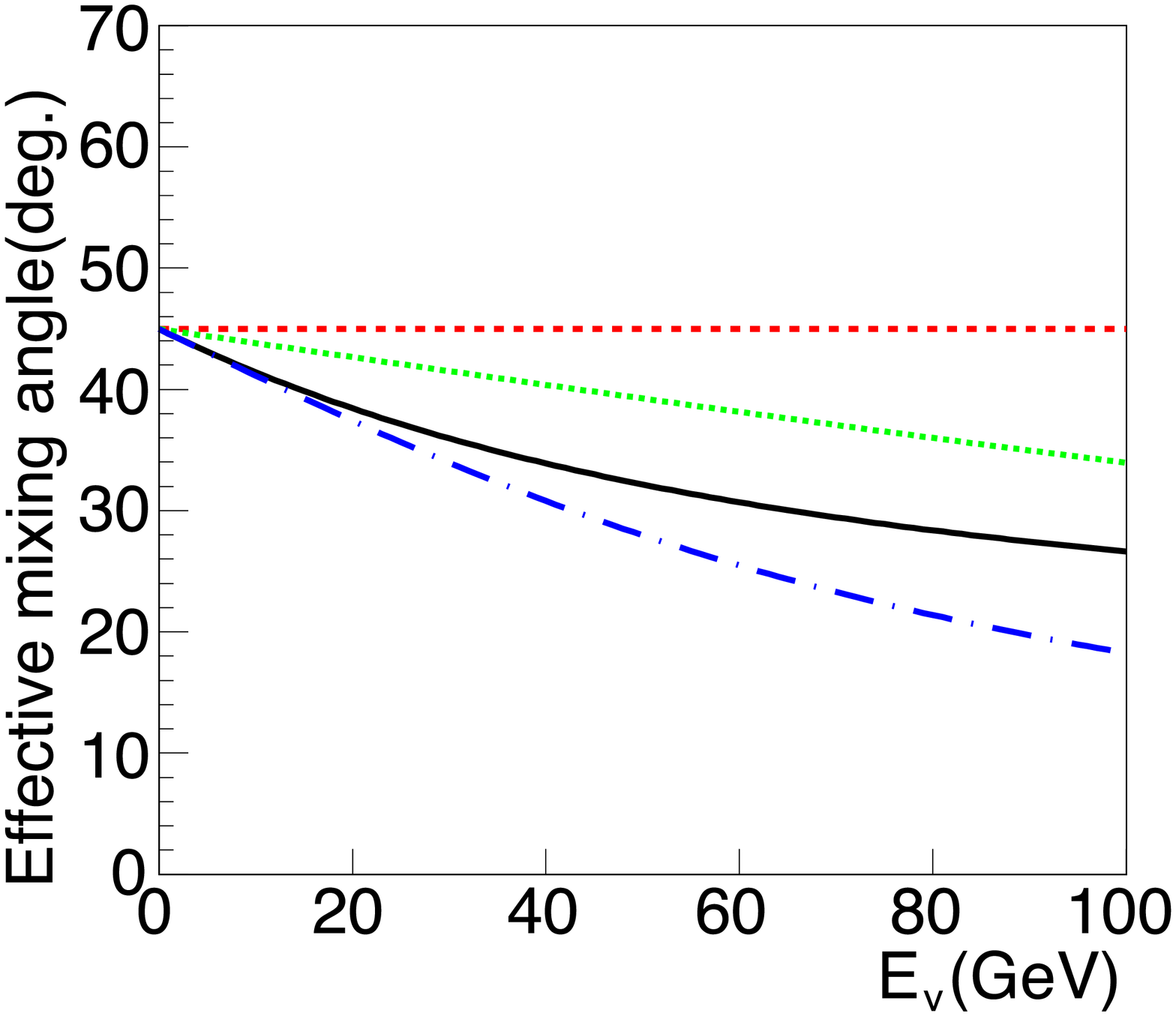} \hspace{0.5cm}
    \includegraphics[width=7.0cm,keepaspectratio]{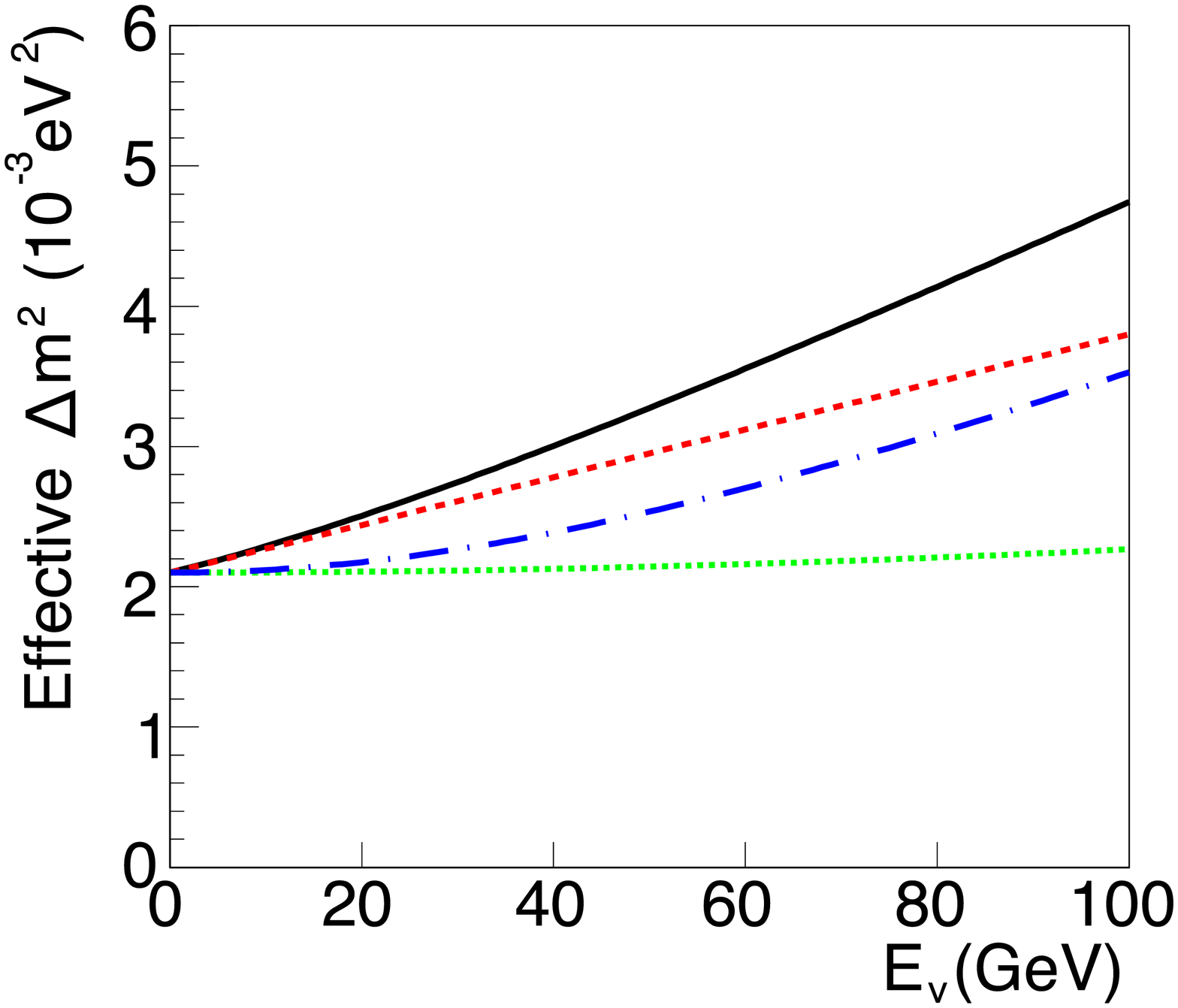}
  \end{center}
  \caption{(Left) Effective mixing angle in matter. (Right) Effective mass squared difference in matter. In both panels, solid black curves indicate the case with ($\varepsilon$ (FCNC), $\varepsilon'$ (NU))$=$(0.015, 0.05), dashed red curves with (0.015, 0.0), dotted green curves with (0.0, 0.015), dashed-dotted blue curves with (0.0, 0.05). As for the vacuum parameters, $\theta_{23}$ = 45$^\circ$ and $\Delta m^2_{23}=2.1\times10^{-3}{\rm eV^2}$ are assumed. Matter density is defined as constant $\rho=5.0$g/cm$^3$.}
  \label{fig:2dnsi_effectiveparams}
\end{figure*}

The 2-flavor hybrid model can be extracted from Eq.(\ref{eq:3dnsi_hamiltonian}) by setting $\Delta m^2_{21} = 0$. 
Following the formalism of M.C. Gonzalez-Garcia and M. Maltoni~\cite{Concha04},
assuming that neutrinos possess non-standard interactions with only $d$-quarks~\cite{Concha99, Fornengo},
and defining  $\emt \equiv \varepsilon$ and  $\ett-\emm \equiv \varepsilon'$
the $\nu_{\mu}$ survival probability in constant density matter is given by
\begin{equation}
  P_{\nu_\mu \to \nu_\mu} = 1-P_{\nu_\mu \to \nu_\tau} = 1-\sin^22\Theta\sin^2\left(\frac{\Delta m^2_{23} L}{4E}R\right).
\end{equation}
\noindent The effective mixing angle, $\Theta$, and the correction factor to the oscillation wavelength, $R$, are given by
\begin{eqnarray}
\sin^22\Theta  &=& \frac{1}{R^2} \left( \mbox{sin}^{2}2\theta + R_{0}^{2} \mbox{sin}^{2} 2 \xi +
                   2 R_{0} \mbox{sin} 2\theta \mbox{sin} 2\xi \right),  \nonumber  \\
R &=& \sqrt{ 1 +  R_{0}^{2} +  2 R_{0} ( \mbox{cos} 2\theta \mbox{cos} 2\xi + \mbox{sin} 2\theta \mbox{sin} 2\xi ) }, \nonumber \\
R_{0} &=& \sqrt{2}G_{F} N_{f} \frac{ 4E }{ \Delta m^{2} } \sqrt{ |\varepsilon |^{2} + \frac{\varepsilon'^{2} }{4} } , \nonumber \\
\xi &=&  \frac{1}{2} \mbox{tan}^{-1} \left( \frac{ 2 \varepsilon }{ \varepsilon' } \right),  
\label{eqn:nsi_eff_mixing}
\end{eqnarray} 
\noindent where $\theta$ is the standard two-flavor mixing angle, and $\xi$ is the NSI-induced effective rotation angle in matter.
Both the effective mixing angle and the correction factor depend on the neutrino energy 
as well as the standard oscillation and NSI parameters. 
These parameters are shown as a function of energy for several values of $\varepsilon$ and $\varepsilon'$ in 
Fig.~\ref{fig:2dnsi_effectiveparams}.

\subsection{Expected Phenomena}\label{chap:2dnsi_expectation}
Because of the energy dependence of the standard oscillation term in Eq.(\ref{eq:3dnsi_hamiltonian}), 
the relative dominance of NSI in the hybrid model is expected to depend on the neutrino energy.
The effects of NSI on neutrino oscillation can be divided into three energy ranges: (1) $E_\nu<$1GeV, (2) 1$<E_\nu<$30GeV, (3) $E_\nu>$30GeV.

\vspace{0.5cm}
(1) $E_\nu<$1GeV

At these low energies, the eigenvalue of the vacuum term $\Delta m^2_{23}/2E_\nu (\gtrsim 1\times 10^{-12}$eV$)$
is larger than that of the NSI matter term $\sqrt{2}G_FN_f\varepsilon (\sim 1 \times 10^{-13}$eV$)$, 
assuming $\Delta m^2_{23} = 2.1\times 10^{-3}{\rm eV^2}$, $N_f \equiv N_d \sim 3N_e$ and $\varepsilon\sim\mathcal{O}(1)$. 
Thus the $\numu \to \nutau$ transition is mostly governed by the standard two-flavor oscillation and there is no 
significant contribution from NSI. 
Note that $\varepsilon\sim\mathcal{O}(1)$ is a conservative assumption according to the NuTeV limit $|\emt| < $ 0.05 \cite{NuTeV_NSI}.

\vspace{0.5cm}
(2) 1$<E_\nu<$30GeV

\begin{figure*}[htbp]
  \begin{center}
    \includegraphics[width=6.5cm,keepaspectratio=true]{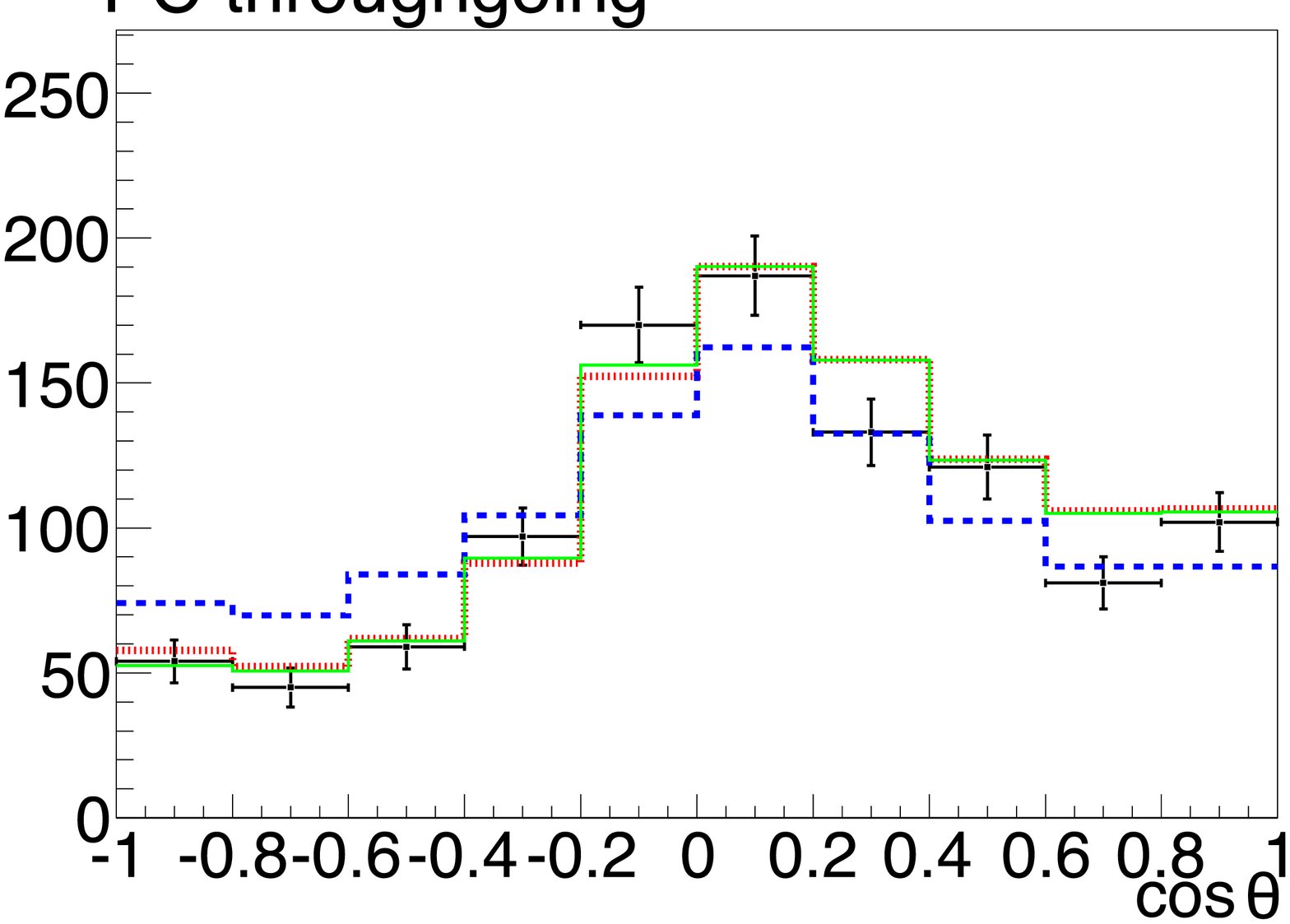} \hspace{1.0cm}
    \includegraphics[width=6.5cm,keepaspectratio=true]{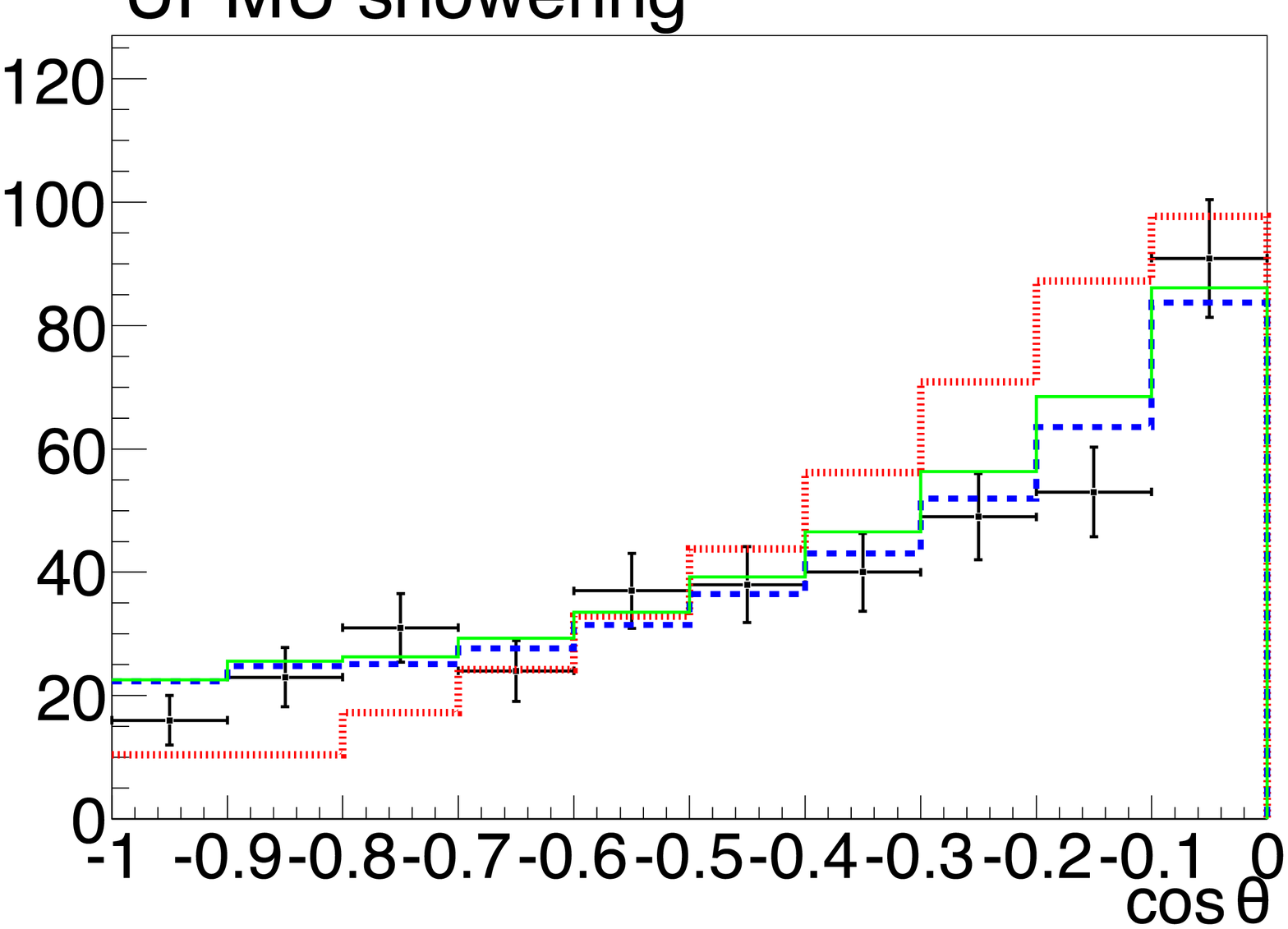}
  \end{center}
  \caption{Zenith angle distribution for PC through-going (left) and UPMU showering (right, included in UPMU through-going where energy loss of muon is caused by pair production, Bremsstrahlung, and photonuclear interactions) sub-samples. In the solid green line ($\varepsilon$, $\varepsilon'$)$=$($1.0\times10^{-3}, -2.4\times10^{-2}$), in the dashed blue line ($1.0\times10^{-3}, -0.38$), and in the dotted red line ($3.2\times10^{-3}, -2.4\times10^{-2}$). In all lines the standard oscillation parameters are $\theta_{23}=45^\circ$ and $\Delta m^2_{23}=2.1\times10^{-3}$eV$^2$.}
  \label{fig:2d_zenith}
\end{figure*}

This energy range corresponds to the FC Multi-GeV (visible energy greater than 1330 MeV), 
PC, and UPMU stopping samples. 
In this region the matter term competes with the vacuum term and thus the $\numu \to \nutau$ transition 
is no longer dominated by standard oscillations but is modified by the matter term. 
The left panel of Fig.~\ref{fig:2dnsi_effectiveparams} shows that for nonzero NU the effective 
mixing angle decreases with increasing neutrino energy, thereby suppressing $\numu \to \nutau$ transitions.
Similarly, the right panel shows that nonzero FCNC interactions affect the frequency of oscillations.
Since the effective mass splitting, $\Delta m_{\textit{eff}}^2 \equiv R \Delta m^{2}_{23}$, is larger than 
$\Delta m^2_{23}$,  the first oscillation maximum is expected to occur 
at higher neutrino energies than for standard oscillations.
Focusing on the zenith angle distributions presented in the left panel of Fig.~\ref{fig:2d_zenith}, 
the magnitude of the $\mu$-like deficit in the upward-going direction is expected to become smaller due to $\varepsilon'$.
Further, the shape of the zenith angle distribution near and above the horizon is modified in the presence of $\varepsilon$ 
for the higher energy $\numu$ samples.

\vspace{0.5cm}
(3) $E_\nu>$30GeV

Above 30GeV, most atmospheric neurinos are in the UPMU through-going sample, which ranges in energy from tens of GeV to $\sim$ 100 TeV.
At these energies vacuum oscillations have less of an effect on the $\numu \to \nutau$ transition, 
while the transitions induced by the matter term become significant when the neutrino path length in the Earth is sufficiently long. 
As shown in the right panel of Fig.~\ref{fig:2d_zenith}, FCNC interactions play a leading role in this energy range 
because the modified oscillation frequency is comparable in size to the oscillation frequency of these neutrinos in the Earth.
Thus, $\numu \to \nutau$ transitions driven by FCNC are expected to occur. 
In contrast, at these energies NU interactions suppress the effective mixing angle 
and therefore are expected to produce a sub-leading effect on the data. 
Note that at the current values of the atmospheric mixing parameters, the standard 
$\numu \to \nutau$ transition is already increasingly supressed at these energies.  
In summary, FCNC interactions induce a faster oscillation frequency and lead to $\numu \to \nutau$ transitions at 
shorter path lengths in matter, while NU suppresses the transition even at these energies.

\subsection{Analysis Method}\label{chap:2dnsi_ana}
We evaluate the agreement between the data and Monte Carlo oscillated
according to the hybrid NSI model using a $\chi^2$ test. 
The SK-I and SK-II data are divided according to their reconstructed event types, momenta, and zenith angles 
into 400 and 350 bins, respectively.
In SK-I (SK-II) there are 310 (280) bins for the FC samples, 60 (40) for the PC samples, and 30 (30) for the upward-going muon samples.
Data from the two run periods are treated separately due to differences in the detector response and in  
the effects on the atmospheric neutrino flux from solar modulations during the runs. 
In order to accurately treat bins with small statistics in this binning scheme, 
a likelihood based on Poisson probabilities~\cite{Pull} is used. 
The complete $\chi^2$ with 750 data bins and 90 systematic uncertainties is defined as
\begin{widetext}
  \begin{equation}
    \chi^2 = 2\sum_{i=1}^{750}
    \left(N_{i}^{\textit{exp}}(1+\sum_{j=1}^{90}f_{j}^{i}\epsilon_{j}) - N_{i}^{\textit{obs}}
      + N_{i}^{\textit{obs}} \ln \frac{N_{i}^{\textit{obs}}}{N_{i}^{\textit{exp}}(1+\sum_{j=1}^{90}f_{j}^{i}\epsilon_{j})} \right)
    + \sum_{j=1}^{90} \left(\frac{\epsilon_{j}}{\sigma^{sys}_j}\right)^2 ,
    \label{equation:chi2def}
  \end{equation}
\end{widetext}
\noindent where $\epsilon_j$ is a fitting parameter for the $j$-th systematic error and $f^i_j$ is the fractional change of the event rate 
in the $i$-th bin due to a $1\sigma$ change in $j$-th systematic error.

Using this equation, a value of $\chi^2$ is evaluated at each point in a four-dimensional parameter space defined by
$\sin^22\theta_{23}$, $\Delta m^2_{23}$, 
log$_{10} \varepsilon$, and log$_{10} \varepsilon'$, where $\varepsilon$ and $\varepsilon'$ range from $1.0 \times 10^{-3}$ to $3.2 \times 10^{-2}$ and from $1.0 \times 10^{-3}$ to $0.42$, respectively.
The fit is performed on a 51$\times$51$\times$51$\times$51 grid in this space. 
In Eq.(\ref{equation:chi2def}), the 90 $\epsilon_j$ parameters are varied to minimize the value of $\chi^2$ 
for each choice of oscillation parameters. 
The point in parameter space returning the smallest value of $\chi^{2}$ is defined as the best fit. 
Since the effects of $\varepsilon$ are symmetric between negative and positive values, we only consider positive $\varepsilon$.

\subsection{Two-Flavor Non-Standard Interaction Analysis Result}\label{chap:2dnsi_result}
The result of a scan on this parameter space gives a best fit at
\begin{align}
  &\sin^{2}2\theta_{23} = 1.00, \hspace{0.3cm} \Delta m^2_{23}=2.2\times10^{-3} {\rm eV}^2,& \nonumber \\
  &~~~\varepsilon = 1.0 \times10^{-3}, \hspace{0.3cm} \varepsilon'=-2.7\times10^{-2}& \nonumber \\
  &~~~\chi^2_{min} = 838.9~/~746.0~{\rm d.o.f.}&
\end{align}

\noindent The best fit value assuming only standard 2-flavor oscillations 
with the same samples and binning is $\chi^2_{min} = 834.3/748$ d.o.f. 
at $\sin^{2}2\theta = 1.00$ and $\Delta m^2 = 2.1\times10^{-3}$eV$^2$. 
%
%
The slightly larger $\chi^2_{min}$ value from the two-flavor hybrid model compared to the standard analysis is caused by the different numerical method used in the NSI analysis:
the chosen ranges of $\varepsilon$ and $\varepsilon'$ do not allow the two-flavor hybrid model to reduce completely to standard oscillations.

\begin{figure}[htbp]
  \begin{center}
    \includegraphics[width=7.cm, keepaspectratio]{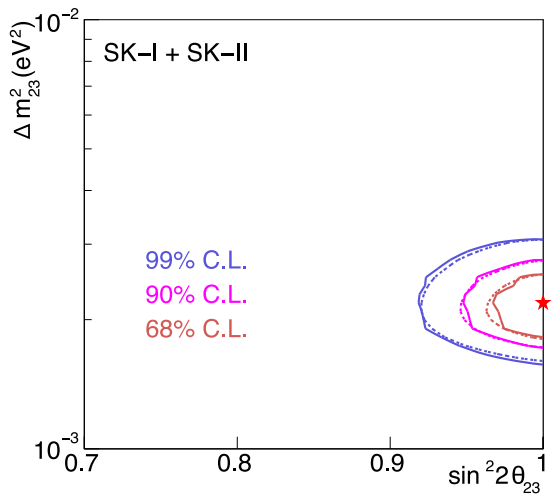}
    \caption{Allowed oscillation parameter regions derived by the 2-flavor hybrid model analysis (solid curves), where undisplayed parameters $\varepsilon$ and $\varepsilon'$ are integrated out. For reference, the result of standard 2-flavor oscillation is added (dashed curves).}
    \label{fig:2dnsi_osc_allowed}
  \end{center}
\end{figure}

Fig.~\ref{fig:2dnsi_osc_allowed} shows the allowed neutrino oscillation parameter regions,
$\sin^22\theta_{23}$ and $\Delta m^2_{23}$, from this analysis and from the standard oscillation analysis.  
The three contours correspond to the 68\%, 90\% and 99\%C.L., defined by $\chi^2 = \chi^2_{min}$ $+$2.30, 4.61, and 9.21, respectively. 
There is no inconsistency between the two sets of allowed regions. 
Since the difference in the minimum $\chi^{2}$ values is not large, 
no significant contribution to standard two-flavor oscillations from NSI effects is found in this analysis.

\begin{figure}[htbp]
  \begin{center}
    \advance\leftskip-1.5cm
    \includegraphics[width=6.cm, keepaspectratio]{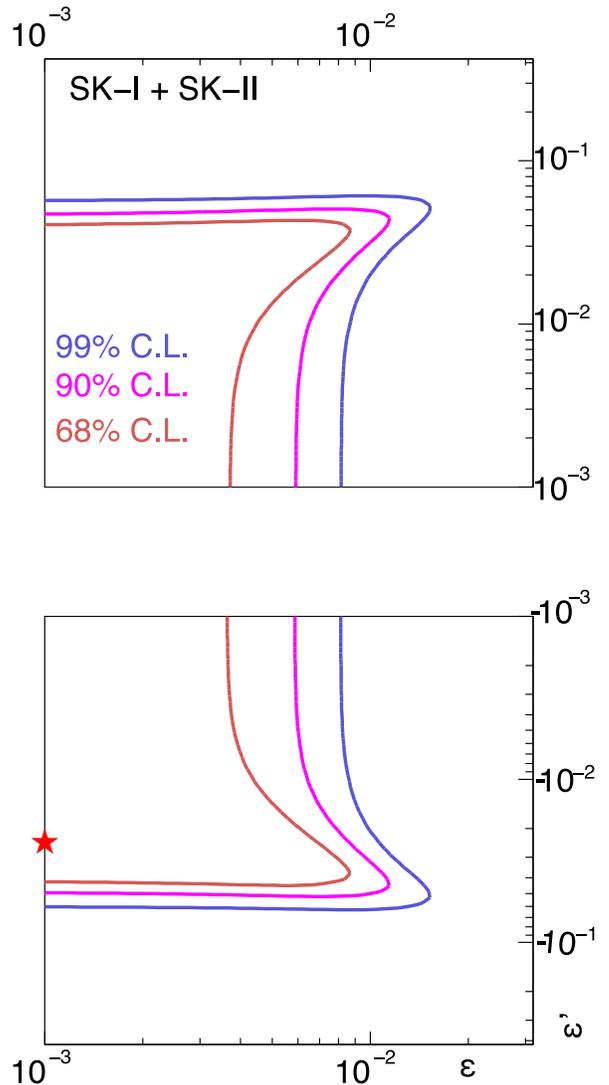}
    \caption{ Allowed NSI parameter regions in the 2-flavor hybrid model. 
              The horizontal axis shows $\varepsilon$ (FCNC) and the vertical axis shows $\varepsilon'$ (NU). 
              Undisplayed parameters $\sin^22\theta_{23}$ and $\Delta m^2_{23}$ are integrated out. 
              The star represents the best-fit point for the NSI parameters.
            }
    \label{fig:2dnsireal_allowed}
  \end{center}
\end{figure}

The allowed regions of the NSI parameters are shown in Fig.~\ref{fig:2dnsireal_allowed}, where there undisplayed parameters $\sin^{2}2\theta_{23}$ and $\Delta m^2_{23}$ have been minimized over. 
At 90\,\% C.L. the obtained limits on the NSI parameters in the $\numu-\nutau$ sector are 
\begin{eqnarray}
  |\varepsilon| < 1.1 \times 10^{-2} \mbox{ and} \hspace{0.2cm} - 4.9 \times 10^{-2} < \varepsilon' < 4.9 \times 10^{-2}.
\end{eqnarray}

These limits can be compared with a phenomenological study using the SK-I (79 kton yr) and MACRO atmospheric neutrino data~\cite{Fornengo}, in which the limits on FCNC and NU at 90\,\% C.L. are $2.0 \times 10^{-2} < \varepsilon < 1.3 \times 10^{-2}$ and $- 4.7 \times 10^{-2} < \varepsilon' < 4.2 \times 10^{-2}$.

\subsection{Discussion}\label{chap:2dnsi_discussion}
\begin{figure}[htbp]
  \begin{center}
    \includegraphics[width=7.cm, keepaspectratio]{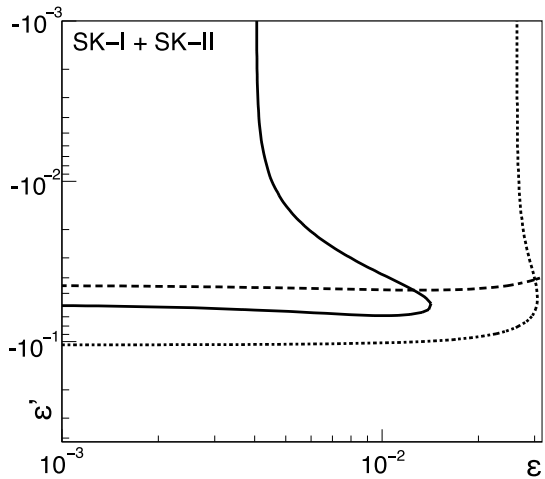}
    \caption{Allowed NSI parameter regions for various event sub-samples. 
             The solid curve indicates the allowed region from the UPMU through-going sample, 
             the dashed curve is that from the PC and UPMU stopping samples, 
             and the dotted curve is from the FC Single-ring Multi-GeV and Multi-ring samples. 
             Each contour corresponds to $\chi^2 = \chi^2 + 2.30$ (68\%C.L.).
            }
    \label{fig:2dnsireal_chi2_sub}
  \end{center}
\end{figure}

As described in Sec.~\ref{chap:2dnsi_expectation}, NSI are expected to affect  
multi-GeV muon neutrinos , and are consequently  constrained by the Super-K $\mu$-like samples. 
Fig.~\ref{fig:2dnsireal_chi2_sub} shows the allowed region spanned by the NSI parameters for three sub-samples of the data.
Since the minimum $\chi^2$ is located in negative $\varepsilon'$ space, the negative $\varepsilon'$ plane is presented. 
The strongest constraint on $\varepsilon$ comes from the UPMU through-going sample (solid curve), 
while $\varepsilon'$ is most tightly constrained by the PC and UPMU stopping (dashed curve) samples.
The constraint from the sub-GeV samples is too weak to be visible in this figure. 

These constraints on $\varepsilon$ can be understood as follows. 
As shown in the right panel of Fig.~\ref{fig:2dnsi_effectiveparams}, FCNC interactions (dashed red curve) increase the effective 
neutrino oscillation frequency thereby shortening the oscillation length for 
neutrinos at a fixed energy. 
Above 30 GeV, where the atmospheric sample is dominated by UPMU events,
the oscillation length in the absence of NSI is already greater than the 
diameter of the Earth. The addition of FCNC effects on the other hand can 
shorten the oscillation length enough to induce oscillations in the steepest 
upward-going bins of the UPMU sample.
This is an apparent contradiction with the data and results in a tight 
constraint on $\varepsilon$. 
Below $\sim 30$ GeV, the oscillation length is sufficiently short 
in both standard and NSI cases to cause oscillations in the FC and PC samples. 
For this reason it is more difficult for these samples to discriminate 
between standard- and FCNC-induced oscillations. 
 
As for the most stringent limit on NU derived from the PC and UPMU stopping samples, since a large $\varepsilon'$ gives a small magnitude of the $\mu$-like deficit in the upward-going direction and a distortion of the shape of the zenith angle distribution near and above the horizon (see dashed blue line in the left panel of Fig.~\ref{fig:2d_zenith}), a tight constraint on NU can occur.
Further limits on NU by other sub-samples can be understood by the discussion in Sec.~\ref{chap:2dnsi_expectation}.

%
%
\section{Analysis with a Three-Flavor Hybrid Model}\label{chap:3d-hybrid}
We now consider a {\it 3-flavor hybrid model} in which NSI in the $\nue-\nutau$ sector 
coexist with standard 2-flavor $\numunutau$ oscillations and all other NSI are zero.
By introducing couplings between $\nue$ and $\nutau$, this model allows flavor transitions of all types. 
That is, an overall  $\numu \to \nue$ transition becomes possible due to the the $\eet$ induced $\nutau \to \nue$
conversion working in conjunction with the standard  $\numunutau$ oscillation: 
\begin{equation}
  \numu \xrightarrow{\theta_{23}} \nutau \xrightarrow{\eet} \nue.
  \label{eq:3dnsi_numu_nue}
\end{equation}
\noindent Note that since the other FCNC epsilons have been set to zero, there is no direct transition between $\nu_{\mu}$ 
and $\nu_{e}$. This analysis therefore aims to constrain possible NSI effects by examining changes in the 
$\nu_{e}$ and $\nu_{\mu}$ fluxes on top of standard oscillation-like effects.

\subsection{Formalism}\label{chap:3dnsi_formalism}
An evolution matrix from time $t_0$ to $t$ can be obtained by diagonalizing the Hamiltonian Eq.~(\ref{eq:3dnsi_hamiltonian}) in terms of the leptonic mixing matrix in matter $U'$ 
and the effective eigenvalues $\hat{H} = {\rm diag}(E_1,E_2,E_3)$. 
In the case of constant matter density, the evolution matrix in natural units is
\begin{equation}
  S_{\beta\alpha}(t,t_0) = \sum_{i=1}^3 (U'_{\alpha i})^\ast U'_{\beta i} {\rm e}^{-iE_i(t-t_0)}, \hspace{0.5cm} \alpha,\beta = e,\mu,\tau.
\end{equation}
Thus the neutrino oscillation probability under the effect of NSI can be expressed as
\begin{equation}
  P_{\alpha\beta} = |S_{\beta\alpha}(t,t_0)|^2.
\end{equation}

\noindent However, in order to account for the varying matter density in the Earth (from 2.5 to 13 g$/$cm$^3$), 
we divide the neutrino path into several constant density steps and calculate the evolution matrix in each. 
The oscillation probability is obtained from the eigenvalues of the product of these matrices.

\subsection{Expected Phenomena}\label{chap:3dnsi_expectation}
As in the 2-flavor hybrid model, the effects of NSI are expected to vary with the neutrino energy. 
We consider three energy ranges: (1) $E_\nu<$1GeV, (2) 1$<E_\nu<$15GeV, and (3) $E_\nu>$15GeV.

\vspace{0.5cm}
(1) $E_\nu<$1GeV

Since the standard oscillation eigenvalue $\Delta m^2/2E_\nu$ is much greater than the matter potential $\sqrt{2}G_FN_d$, 
$\numu \to \nutau$ transitions induced by these oscillations are expected to be dominant
 and the effects from NSI can be ignored.

\vspace{0.5cm}
(2) 1$<E_\nu<$15GeV

In this energy range, the matter term has a sizable effect on the $\numu \to \nutau$ transition as it did in the 2-flavor hybrid model. 
Moreover, since the current limits on the NSI parameters governing $\nue \to \nutau$ transitions 
are poor: $\eet \sim\mathcal{O}(10^{-1})$  \cite{NSI_limit}, large effects from NSI are possible.

\begin{figure*}[htbp]
  \begin{center}
    \includegraphics[width=6.5cm,keepaspectratio=true]{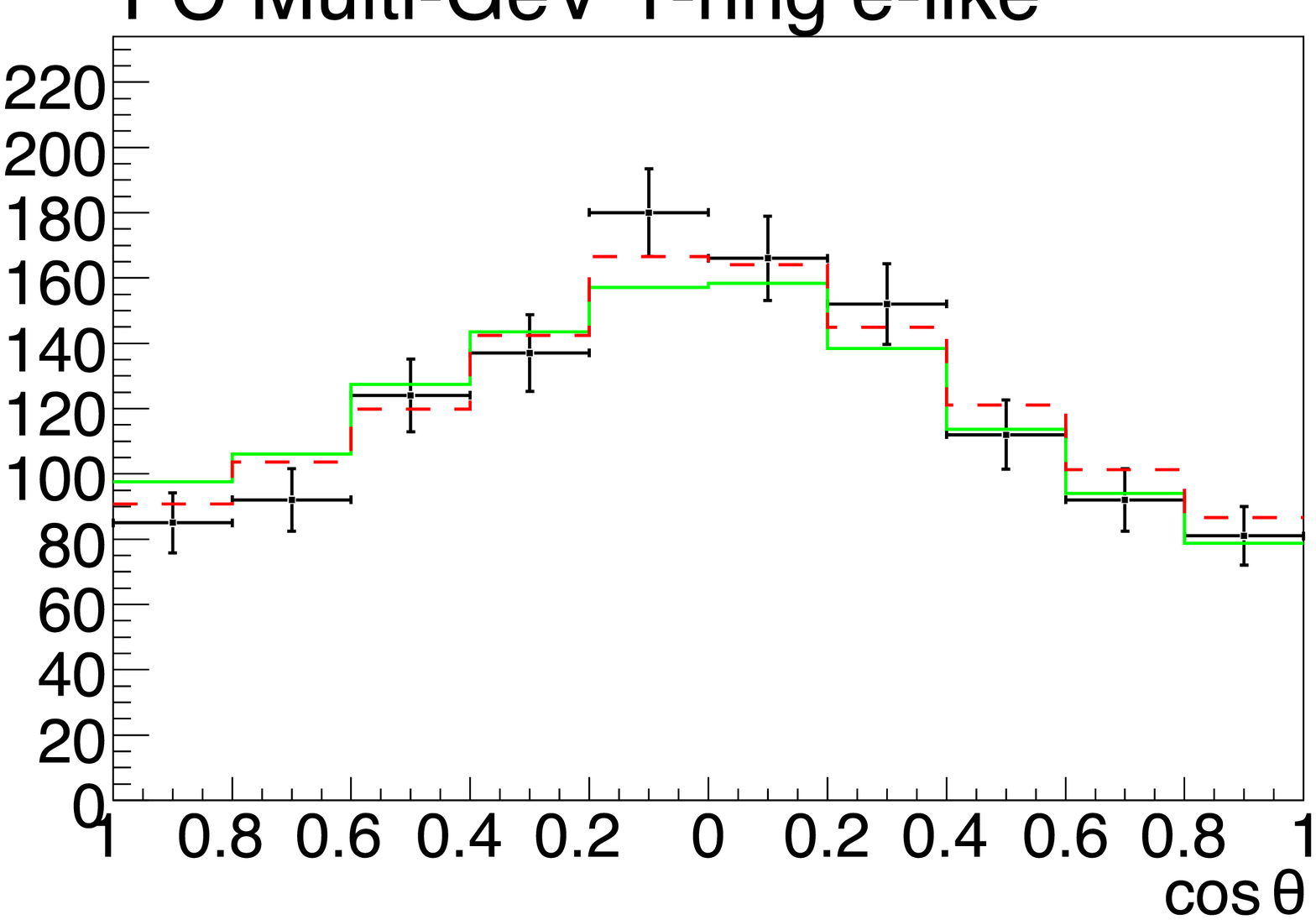} \hspace{1.0cm}
    \includegraphics[width=6.5cm,keepaspectratio=true]{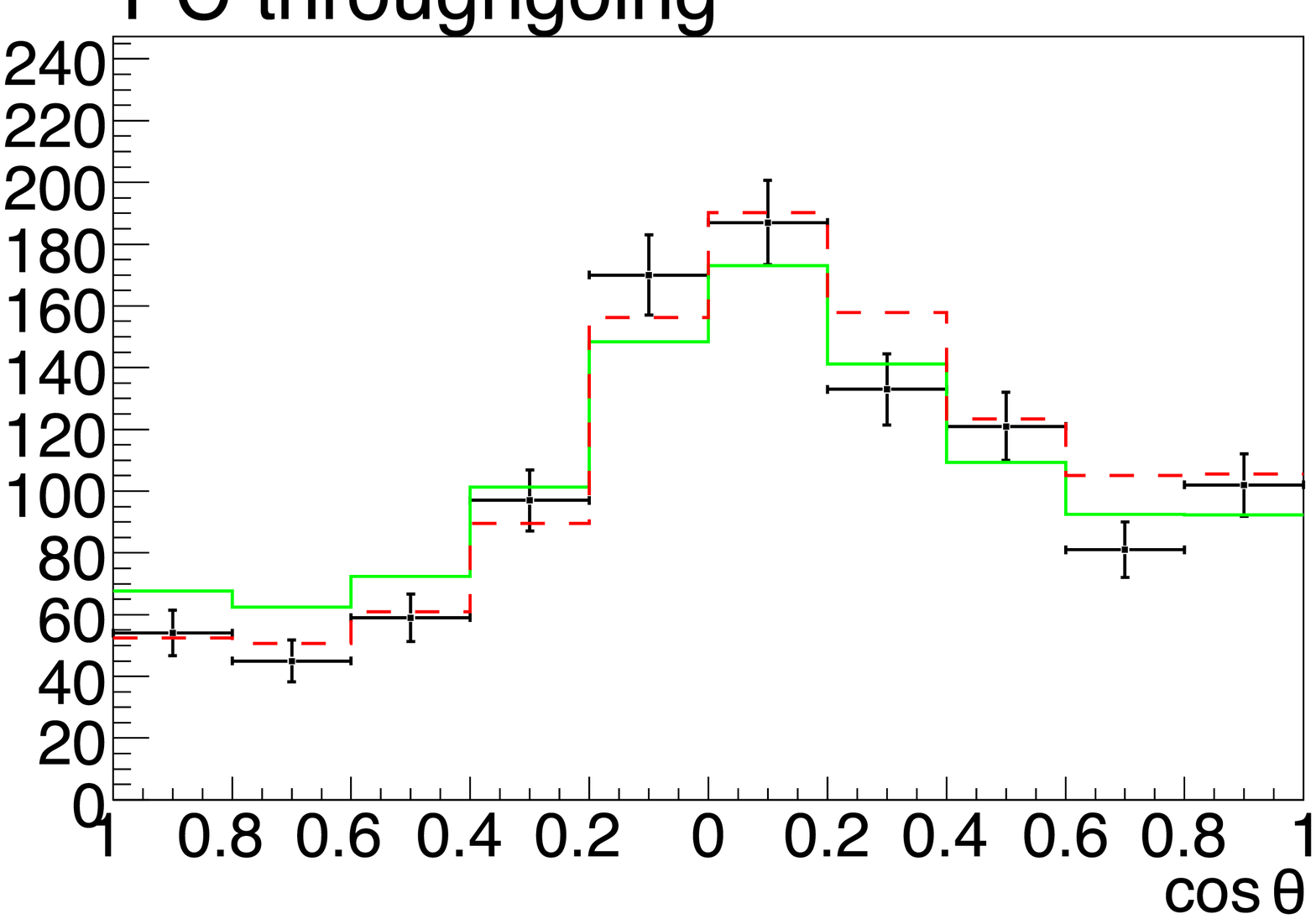}
  \end{center}
  \caption{Zenith angle distribution for typical $e$-like (left, FC Multi-GeV 1-ring $e$-like) and $\mu$-like (right, PC throughgoing) samples. The solid green line is the MC prediction at $\eee=0.0$, $\eet=0.2$, and $\ett=0.2$. The dashed red line is that for standard neutrino oscillations. In all lines the standard oscillation parameters are $\theta_{23}=45^\circ$ and $\Delta m^2_{23}=2.1\times10^{-3}$eV$^2$.}
  \label{fig:3d_zenith}
\end{figure*}

The modification of the $\nue$ flux by NSI can be parameterized by two oscillation probabilities:
$P(\nue \to \nue)$ and $P(\numu \to \nue)$ in Eq.~(\ref{eq:3dnsi_numu_nue}).
Normalizing by the $\nue$ flux$(\Phi_e)$, and disregarding any $\numu$ contamination, 
the resulting $e$-like distributions can be approximately expressed as 
$P(\nue \to \nue)  +  r P(\numu \to \nue)$, 
where $r \equiv \Phi_\mu/\Phi_e$ is the neutrino flavor~ratio \cite{flux:honda2001}.
This ratio grows with increasing neutrino energy and though it is highly dependent on the 
neutrino zenith angle, near the horizon it is $\sim$ 2 up to around 10 GeV. 
Thus, if $P(\nue \to \nue) + r P(\numu \to \nue) \lesssim 1$, as is the case when $\eet$ and $\ett$ have 
comparable values, then since $P(\nue \to \nue)$ is suppressed by $\eet$ and $P(\numu \to \nue)$ by $\ett$, 
the number of $e$-like events at the horizon can be expected to decrease.
 In contrast, the number of $e$-like events in the upward-going bins would effectively increase
because large values of $\eet$ produce additional $\nue$ from $\nutau$ created during standard oscillations. 
This effect is enhanced by the larger flavor ratio in these regions and enables the multi-GeV 
$e$-like samples to help constrain $\eet$. These effects are shown in the left panel of Fig.~\ref{fig:3d_zenith}.

Effects driven by $\ett$ on the other hand may be described in terms of the limiting case where the other NSI are set to zero.
In this case the problem reduces to the 2-flavor hybrid model with $\varepsilon =0$ and $\varepsilon' = \ett$. 
Therefore a constraint on $\epsilon'$ similar to that from the 2-flavor case can be expected.

\vspace{0.5cm}
(3) $E_\nu>$15GeV

  Above a few tens of GeV the $\nue$ flux decreases as their parent muons increasingly reach the ground before decaying, 
making it possible to neglect the $\nu_e$ contribution to the $\numu$ flux from oscillations induced by $\eet$.
Conversely, changes in that flux can be clearly recognized as NSI-driven $\nutau \to \nue$ conversion at these energies.
As in the 2-flavor hybrid case, $\ett$ can be expected to suppress the effective mixing angle and 
increase the effective mass splitting in the $\numu \to \nutau$ sector. Both of these effects can be seen 
in the right panel of Fig.~\ref{fig:3d_zenith}. 

\vspace{0.5cm}
As pointed out in Ref.~\cite{lunardini} the atmospheric neutrino sample cannot effectively constrain 
$\eee$. This is due to the fact that when $\eet$ is zero, 
the eigenstates of the Hamiltonian in matter are identical to the vacuum eigenstates. 
In this case, the matter eigenvalues are no longer dependent upon 
$\eee$ and the problem reduces to two-flavor NSI mixing with 
$\nu_{\mu} \leftrightarrow \nu_{\tau}$ transitions in matter modified by only $\ett.$ 
Despite the lack of sensitivity to $\eee$, at fixed non-zero values it dictates 
a parabolic relationship among the NSI parameters, which in the limit where one of the matter eigenvalues is 
small takes the form  
\begin{equation}
\ett \sim \frac{3|\eet|^{2} }{1 + 3\eee}.
\end{equation}
\noindent This is a general feature which will present itself in the shape of our allowed regions below. 
 
\subsection{Analysis Method}\label{chap:3dnsi_ana}
The analysis procedure for the 3-flavor hybrid model follows that used for the 2-flavor hybrid model.
A value of $\chi^2$ is evaluated at each grid point in the three-dimensional parameter space of $\eee$, $\eet$ and $\ett$, where 51 points are chosen for each.
Since $\eet$ enters the oscillation equations as $|\eet|$~\cite{lunardini} when 
$\theta_{13}$  and solar oscillations ($\Delta m^2_{12}$ and $\theta_{12}$) are not considered,
we only test positive values of the parameter. 

We first set the standard 2-flavor parameters to ($\sin^2\theta_{23}, \Delta m^2$)=(0.5, 2.1$\times$10$^{-3}$eV$^2$)
as motivated by the results from the 2-flavor hybrid model and other Super-K analyses~\cite{sk3flavor}.
Here the standard oscillation parameters are taken as fixed values. 
This assumption will be verified later by comparing the allowed NSI parameter regions derived 
using slightly different standard oscillation parameters.
Modifications to the fitting results when $\theta_{13}$ is non-zero are presented in Sec.~\ref{chap:appendix}

\subsection{Results of the 3-Flavor NSI analysis}\label{chap:3dnsi_result}
\begin{figure*}[htbp]
  \begin{center}
    \centering
    \subfigure[$\eee=-0.50$]{\epsfig{figure=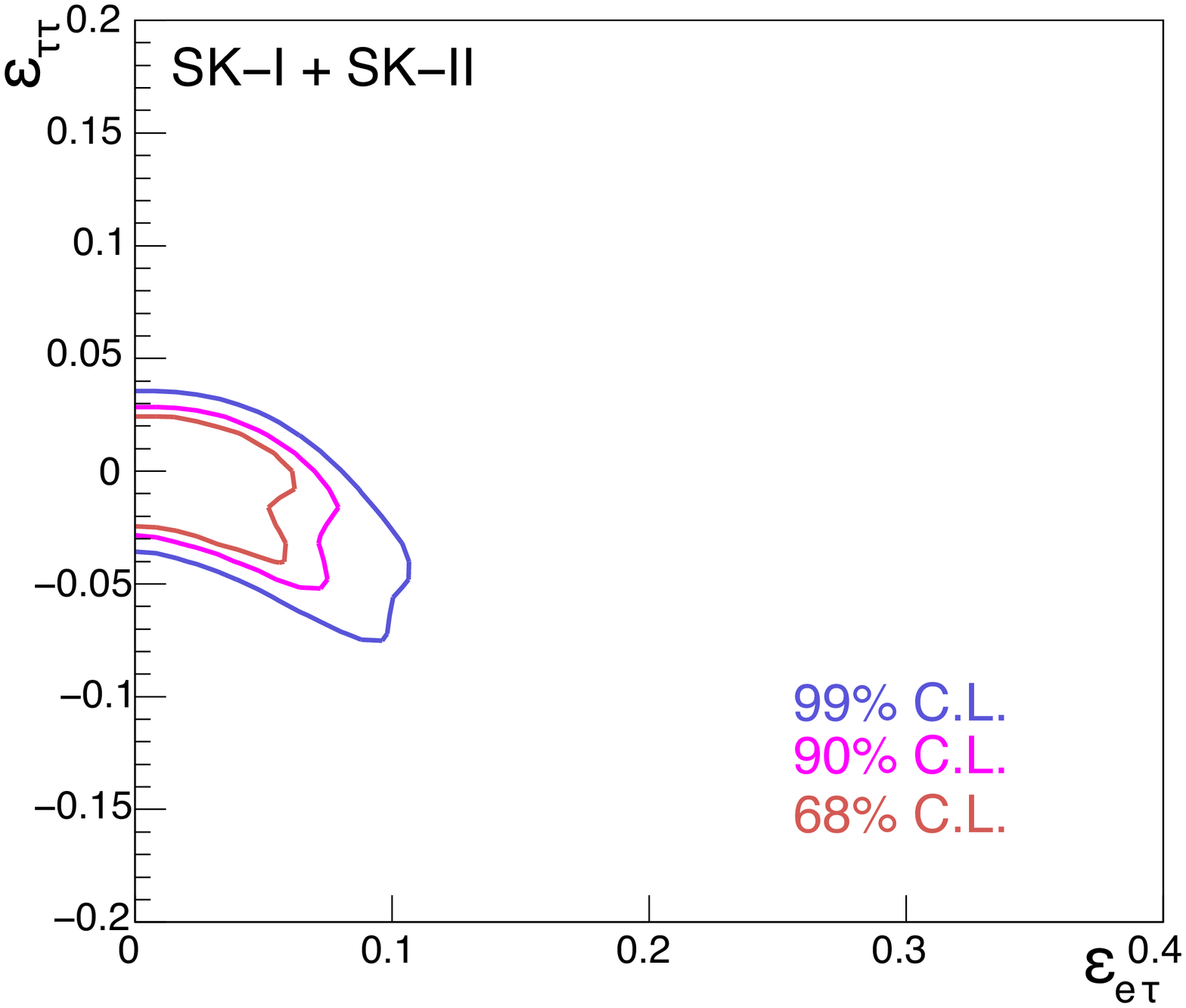,width=.30\textwidth,clip=}}
    \subfigure[$\eee=-0.25$]{\epsfig{figure=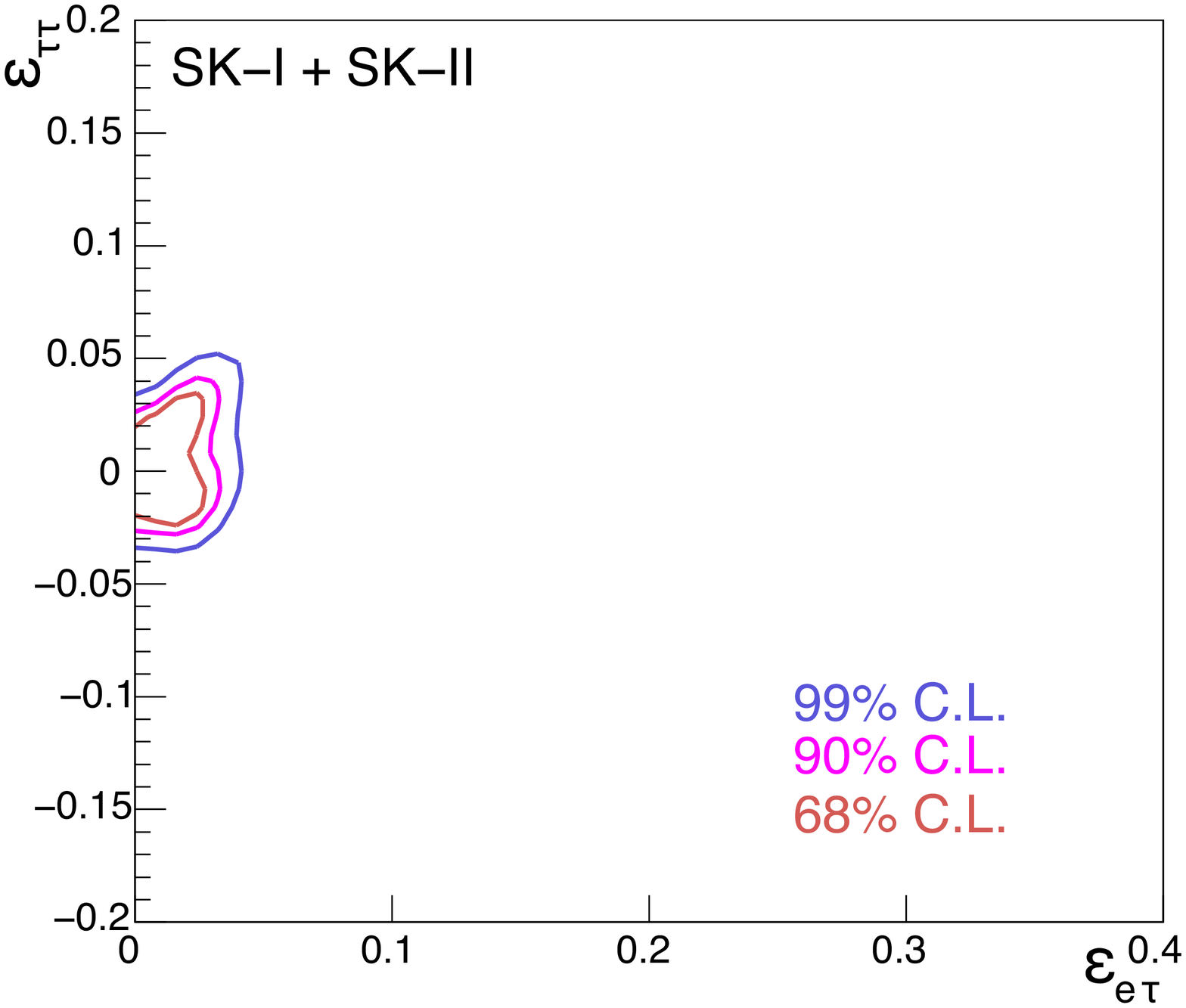,width=.30\textwidth,clip=}}
    \subfigure[$\eee=0.00$]{\epsfig{figure=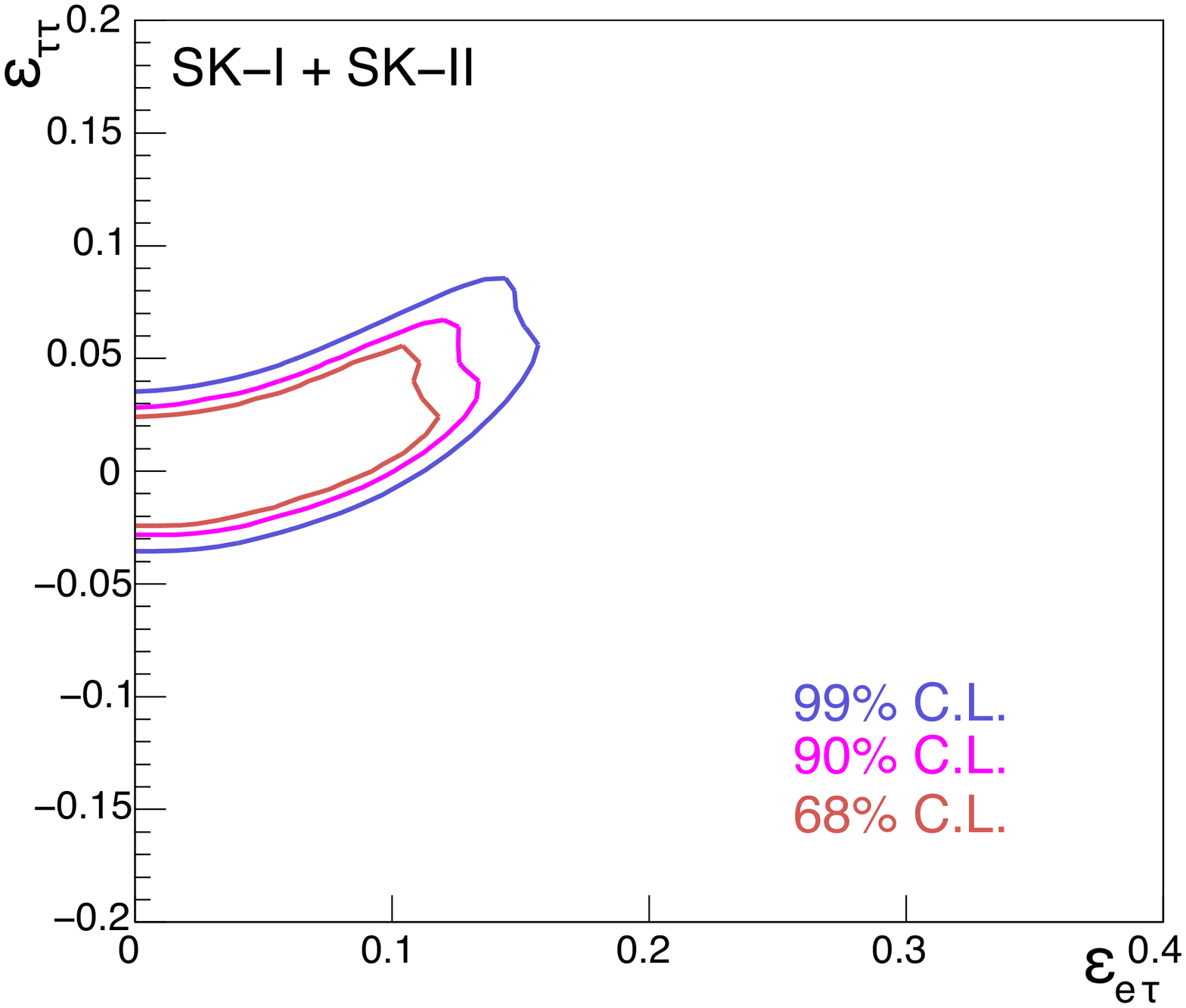,width=.30\textwidth,clip=}} \\
    \subfigure[$\eee=0.25$]{\epsfig{figure=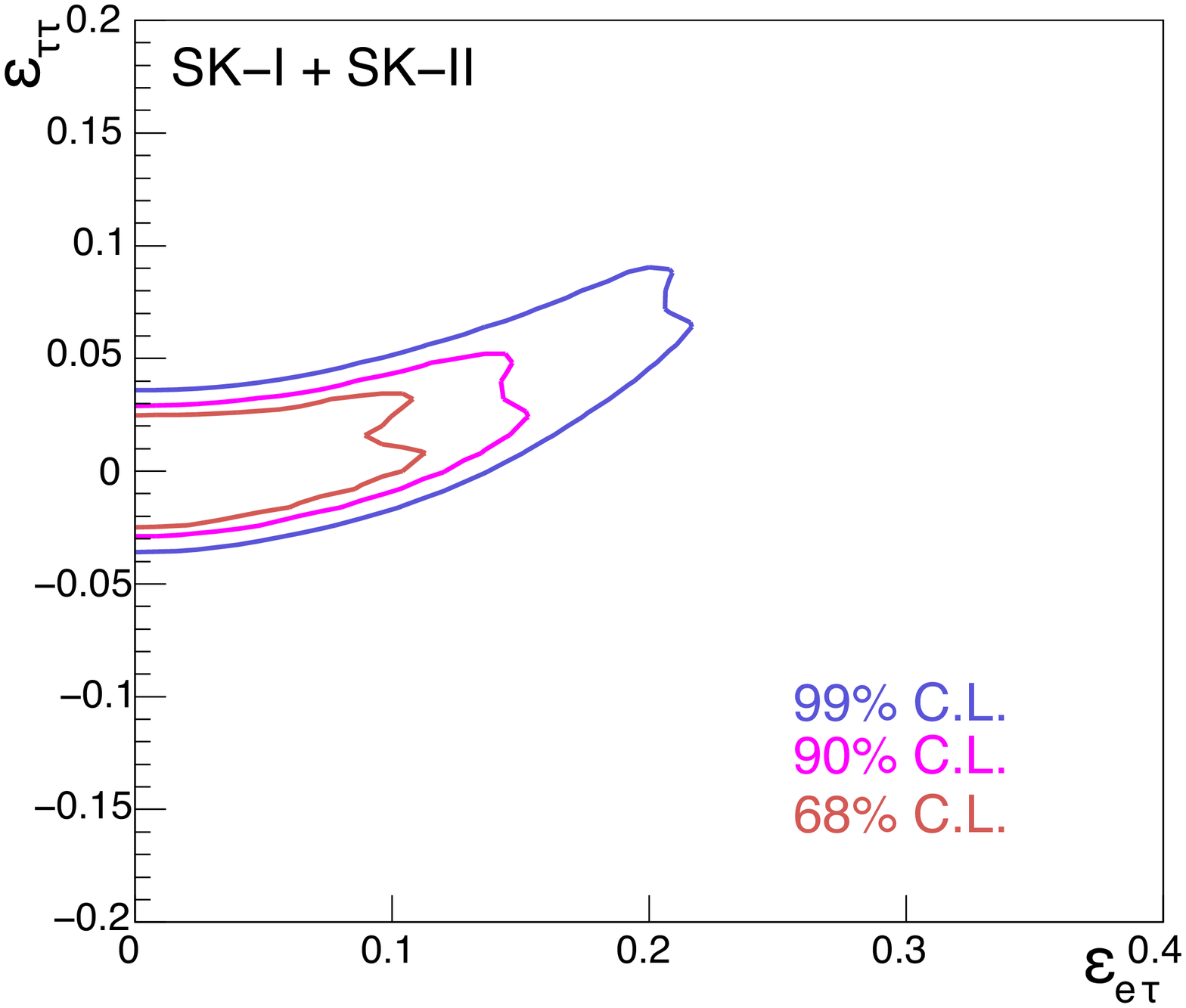,width=.30\textwidth,clip=}}
    \subfigure[$\eee=0.50$]{\epsfig{figure=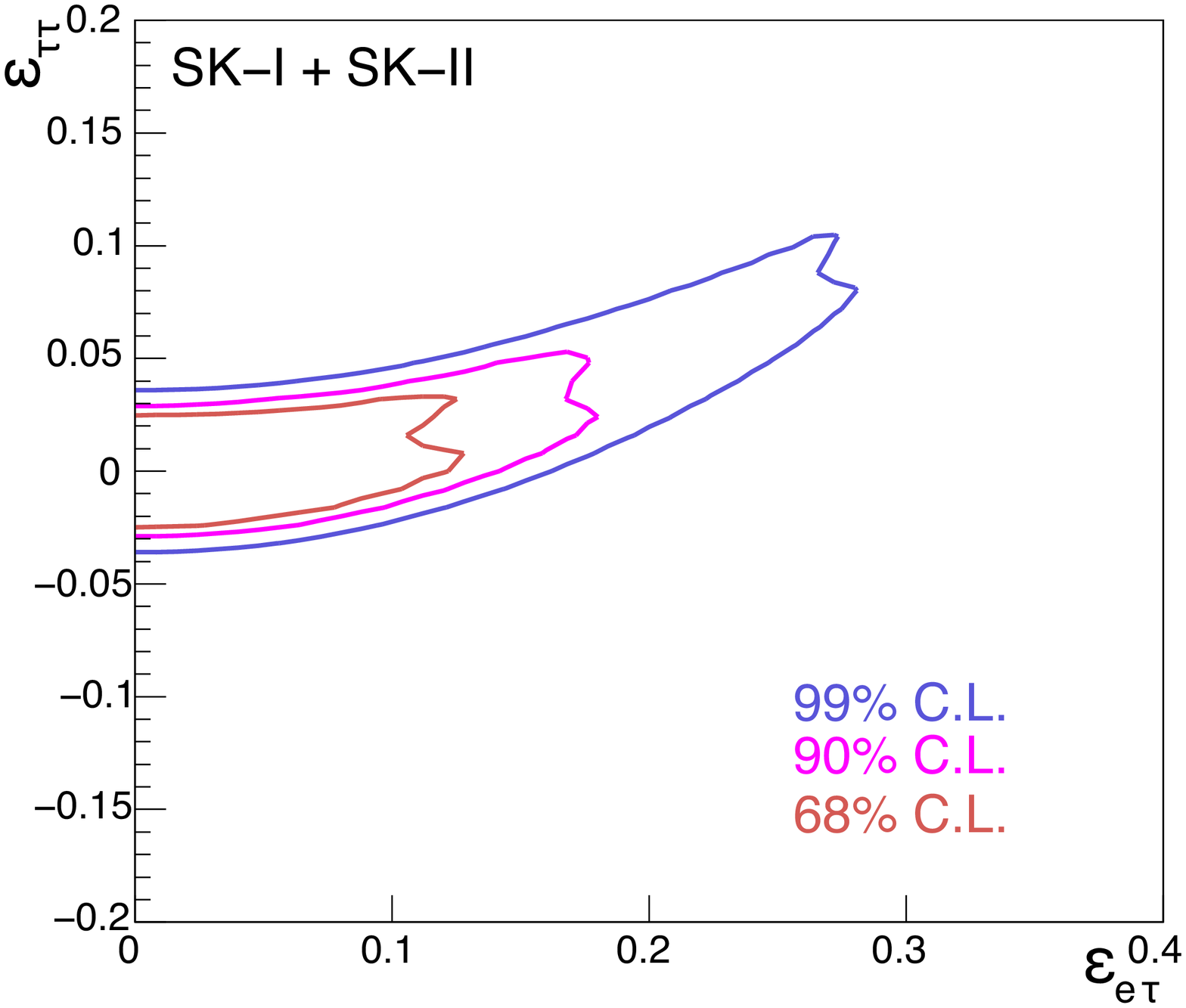,width=.30\textwidth,clip=}}
  \end{center}
  \caption{
           Allowed NSI parameter regions at fixed $\eee$ at 68\%, 90\% and 99\% C.L., where 
           contours are drawn at $\chi^{2} - \chi^{2}_{min} = 2.30, 4.61$, and 9.21, respectively. 
          }
  \label{fig:3dnsi_allowed_main_fixed_eee}
\end{figure*}

The allowed regions for the 3-flavor NSI parameters are shown in Fig.~\ref{fig:3dnsi_allowed_main_fixed_eee}.
Allowed regions for five fixed values of $\eee$ (-0.50, -0.25, 0.00, 0.25, and 0.50) are presented. 
The three contours correspond to the 68\%, 90\% and 99\% C.L. regions as defined by 
$\chi^2 = \chi^2_{min}$ $+$2.30, 4.61, and 9.21, respectively. 
The best-fit values for each $\eee$ are summarized in Tbl.~\ref{3dnsi_bestfittable}.

\begin{table}
  \begin{center}
    \begin{tabular}{r|c|c|c}
      \hline \hline
      $\eee$ & Best-fit $\eet$ & Best-fit $\ett$ & Minimum $\chi^2$ \\
      \hline
      -0.50 & 0.016 & -0.016 & 831.1 \\
      \hline
      -0.25 & 0.016 &  0.024 & 829.9 \\
      \hline 
      0.00  & 0.024 &  0.016 & 830.9 \\
      \hline
      0.25  & 0.000 & -0.016 & 831.4 \\
      \hline
      0.50  & 0.000 & -0.016 & 831.4 \\
      \hline
    \end{tabular}
    \caption{
             The allowed NSI parameters at the 90$\%$ C.L. as a function of $\eee$. 
            }
    \label{3dnsi_bestfittable}
  \end{center}
\end{table}

In these figures, except for the projection for $\eee = -0.25$, parabola-like regions in the $\eet$ - $\ett$ are shown.
In particular, the allowed region extends to negative $\ett$ values when $\eee = -0.5$, 
while they extend to the positive values above $\eee = 0$. 
This feature is consistent with a transition of the matter eigenvalue hierarchy discussed in Ref.~\cite{lunardini}.

\begin{figure*}[htbp]
  \begin{center}
    \centering
    \subfigure[$\eee=-0.50$]{\epsfig{figure=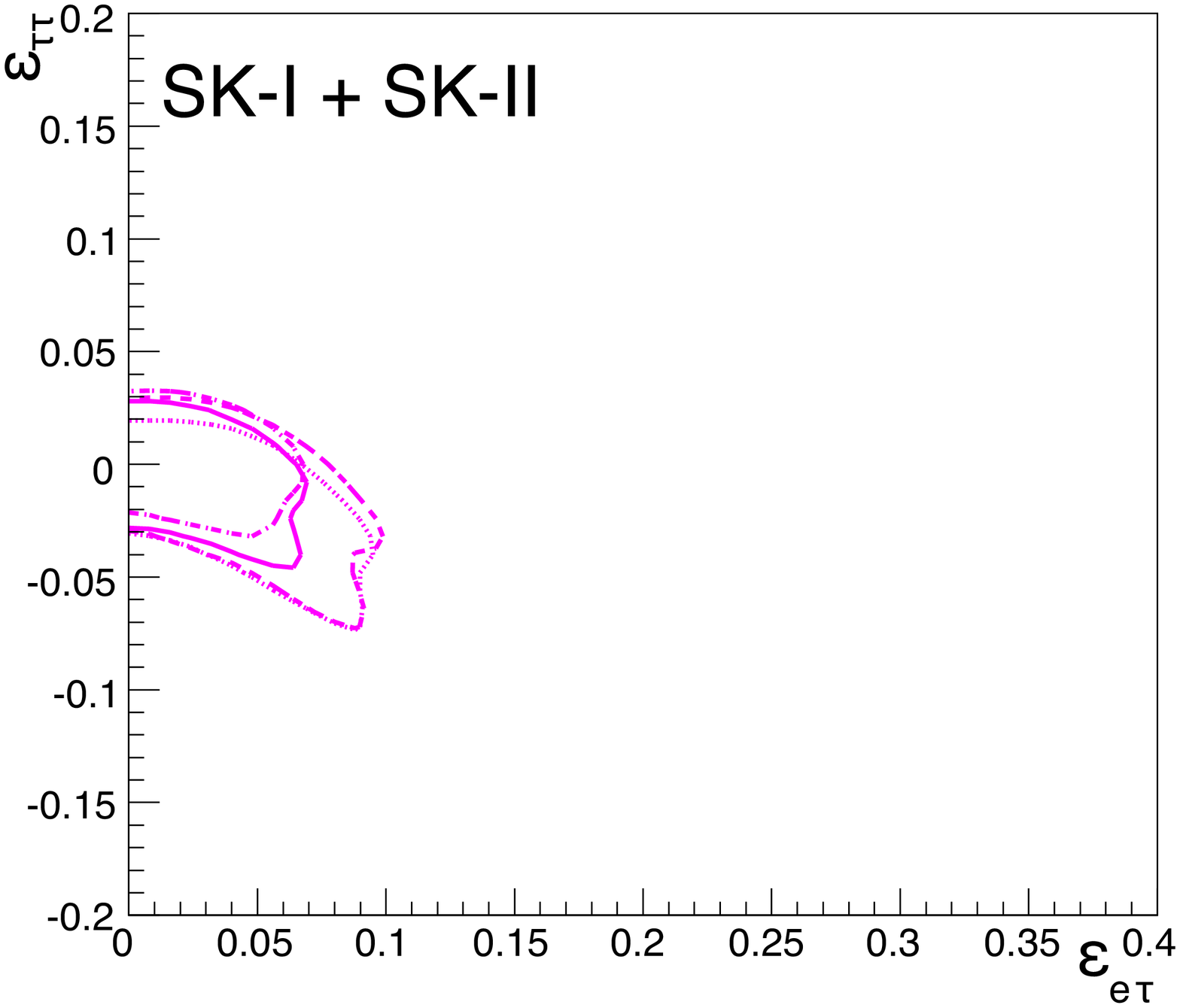,width=.30\textwidth,clip=}}
    \subfigure[$\eee=0.00$]{\epsfig{figure=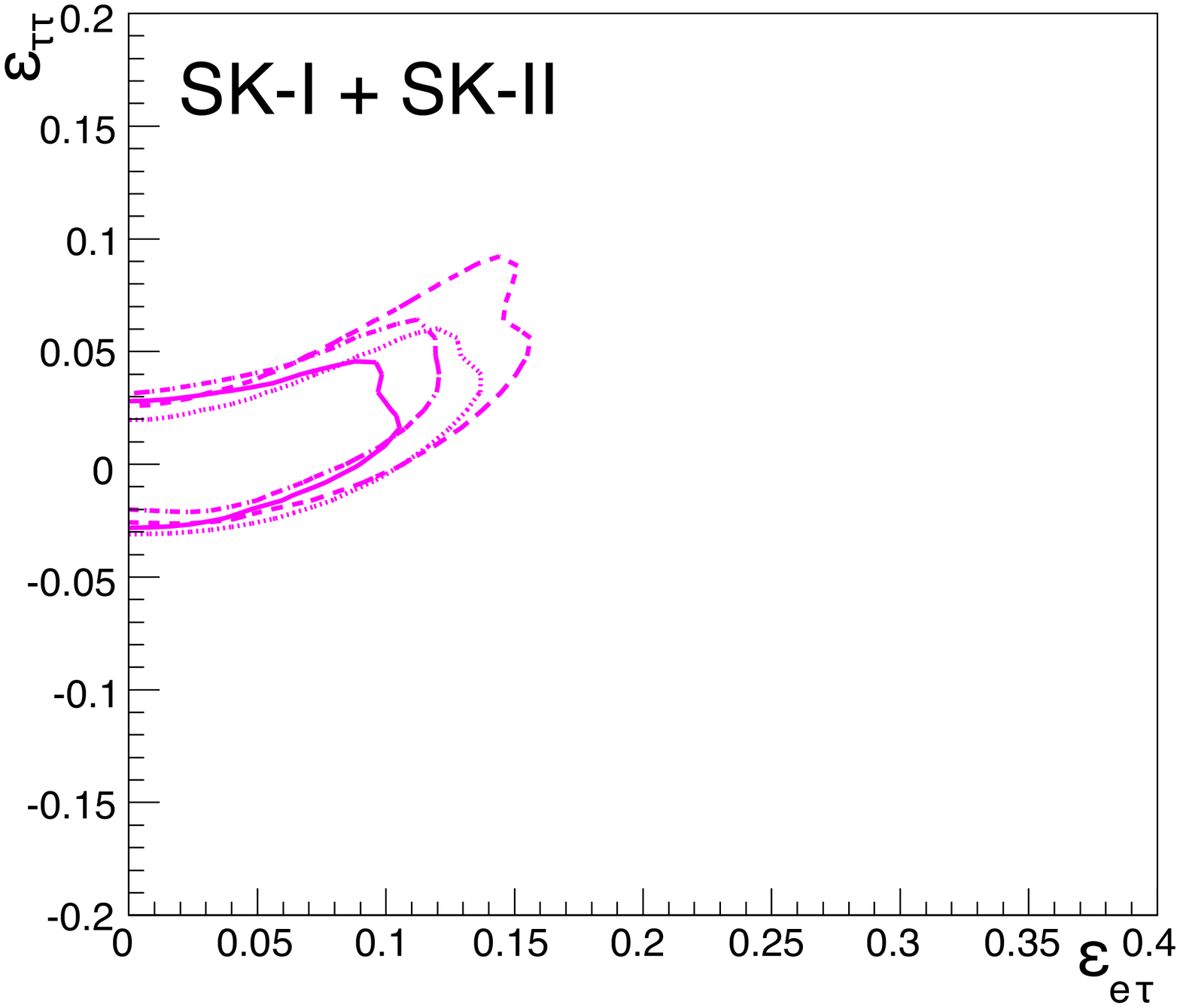,width=.30\textwidth,clip=}}
    \subfigure[$\eee=0.50$]{\epsfig{figure=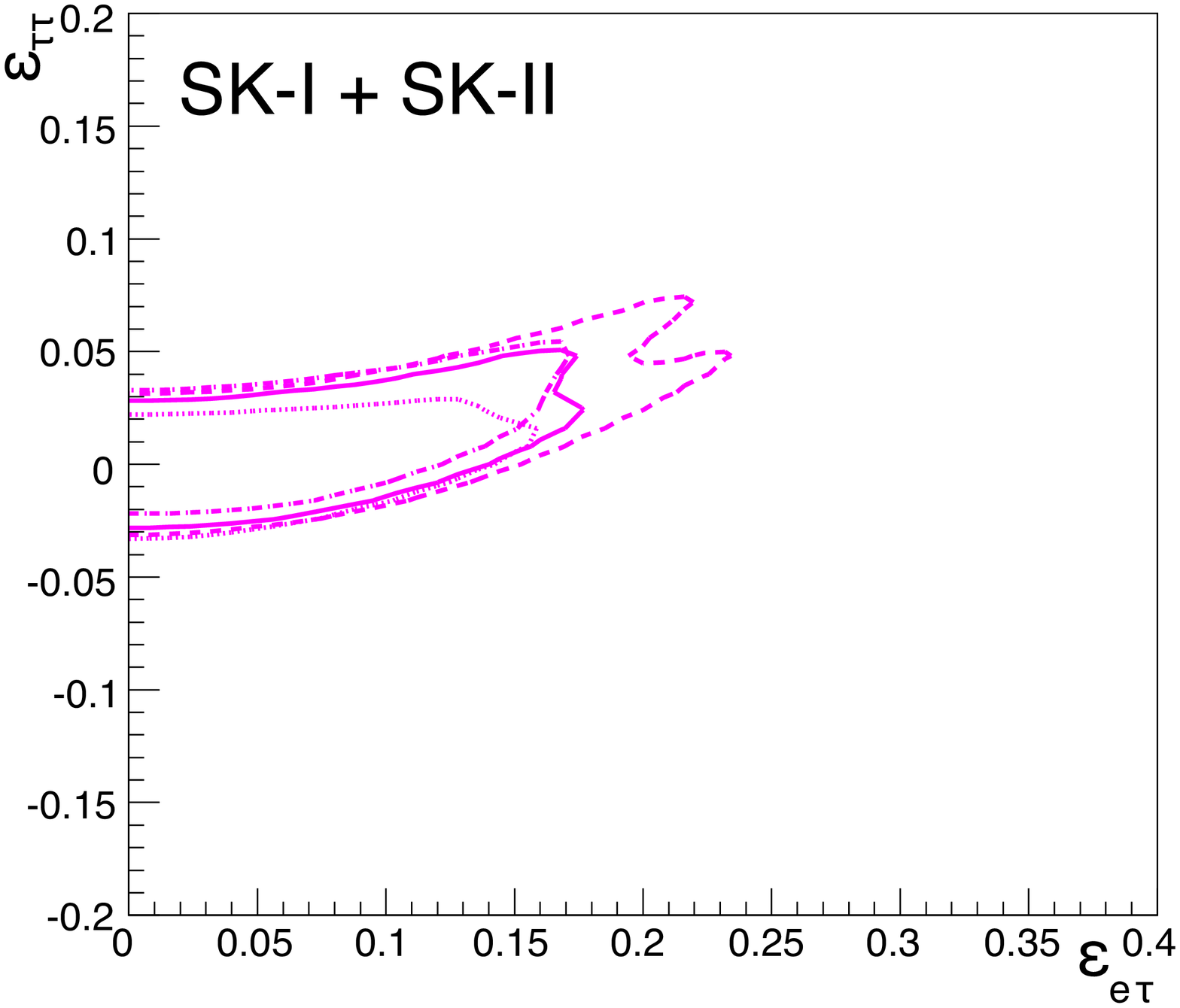,width=.30\textwidth,clip=}}
  \end{center}
  \caption{
             Allowed NSI parameter regions in the $\eet$ vs. $\ett$ plane  
             at the 90$\%$ C.L. for 
             $\eee$ = -0.5, 0.0, and 0.5 using four sets of standard oscillation parameters. 
             The solid curve corresponds to 
             ($\sin^2\theta_{23}, \Delta m^2$)=(0.5, 1.7$\times$10$^{-3}$eV$^2$), 
             the dashed curve to (0.5, 2.7$\times$10$^{-3}$eV$^2$), 
             the dotted curve to (0.39, 2.1$\times$10$^{-3}$eV$^2$), 
             and the dashed-dotted curve to (0.61, 2.1$\times$10$^{-3}$eV$^2$).
           }
  \label{fig:3dnsi_allowed_verify}
\end{figure*}

Next, we evaluate the influence of using fixed standard oscillation parameters. 
Fig.~\ref{fig:3dnsi_allowed_verify} shows the allowed NSI parameter regions 
using four slightly different sets of standard oscillation parameters. 
The oscillation parameters are taken from the 90\% C.L. allowed region from the standard SK 2-flavor analysis.
Although slight changes appear when the NSI fits are repeated using these parameters, 
their minimum $\chi^2$ values are larger than that of the original fit by less than 1.5 units. 
Therefore no significant change in the allowed parameter regions is expected even if 
the standard oscillation parameters are allowed to vary during the NSI fit. 
Finally note that the current allowed regions of $\sin^2\theta_{23}$ and $\Delta m^2$ 
are constrained more tightly by analyses using the full SK data set, which is larger than the sample used here. 

\subsection{Discussion}\label{chap:3dnsi_discussion}
\begin{figure}[htbp]
  \begin{center}
    \includegraphics[width=7.cm, keepaspectratio]{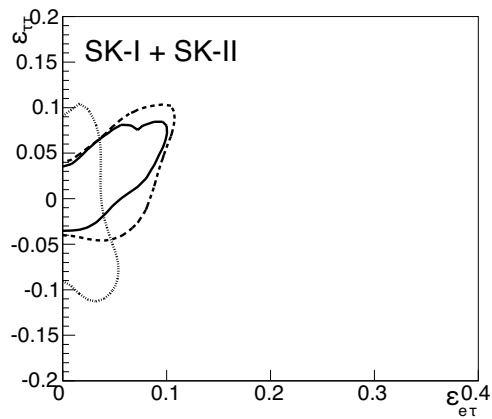}
    \caption{ Allowed NSI parameter regions at 90\% C.L. derived for three subsamples at $\eee = -0.25$. 
              The solid curve indicates the allowed region given by the UPMU throughgoing sample,
              the dashed curve shows the PC and UPMU stopping sample, and the dotted curve 
              is the constraint from the FC multi-GeV samples. 
            }
    \label{fig:3dnsi_allowed_main-subsamples}
  \end{center}
\end{figure}

Fig.~\ref{fig:3dnsi_allowed_main-subsamples} shows the allowed parameter regions for three subsets of 
the atmospheric sample. 
The high energy $\numu$-rich UPMU throughgoing sample (indicated by the solid curve) and PC+UPMU stopping samples 
constrain $\ett$ as expected. 
However, these samples provide no significant contribution to the
constraint on $\eet$ because they lack an $e$-like component. 
On the other hand, the FC multi-GeV samples (shown by the dotted curve) include several $e$-like sub-samples 
which better constrain $\eet$.

%
%
\section{Analysis with a Three-Flavor Hybrid Model with nonzero $\theta_{13}$}\label{chap:appendix}
There are several scenarios in which sub-dominant effects from 
standard neutrino oscillations may affect the allowed NSI 
parameters. In this section we consider how our limits 
change when effects from nonzero $\theta_{13}$ are included 
in the analysis.

\begin{figure*}[htbp]
  \begin{center}
    \centering
    \subfigure[$\eee=-0.50$]{\epsfig{figure=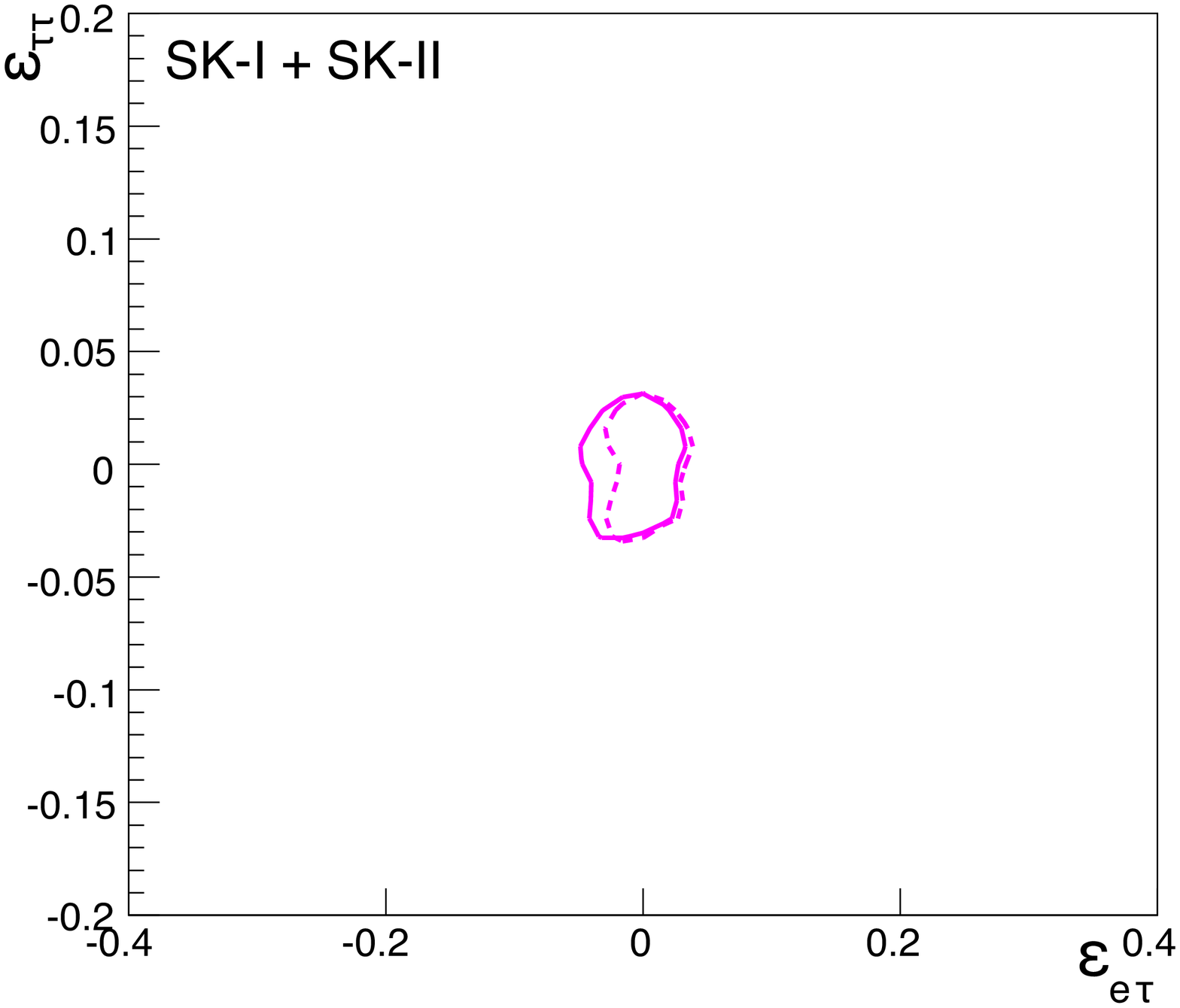,width=.30\textwidth,clip=}}
    \subfigure[$\eee=-0.25$]{\epsfig{figure=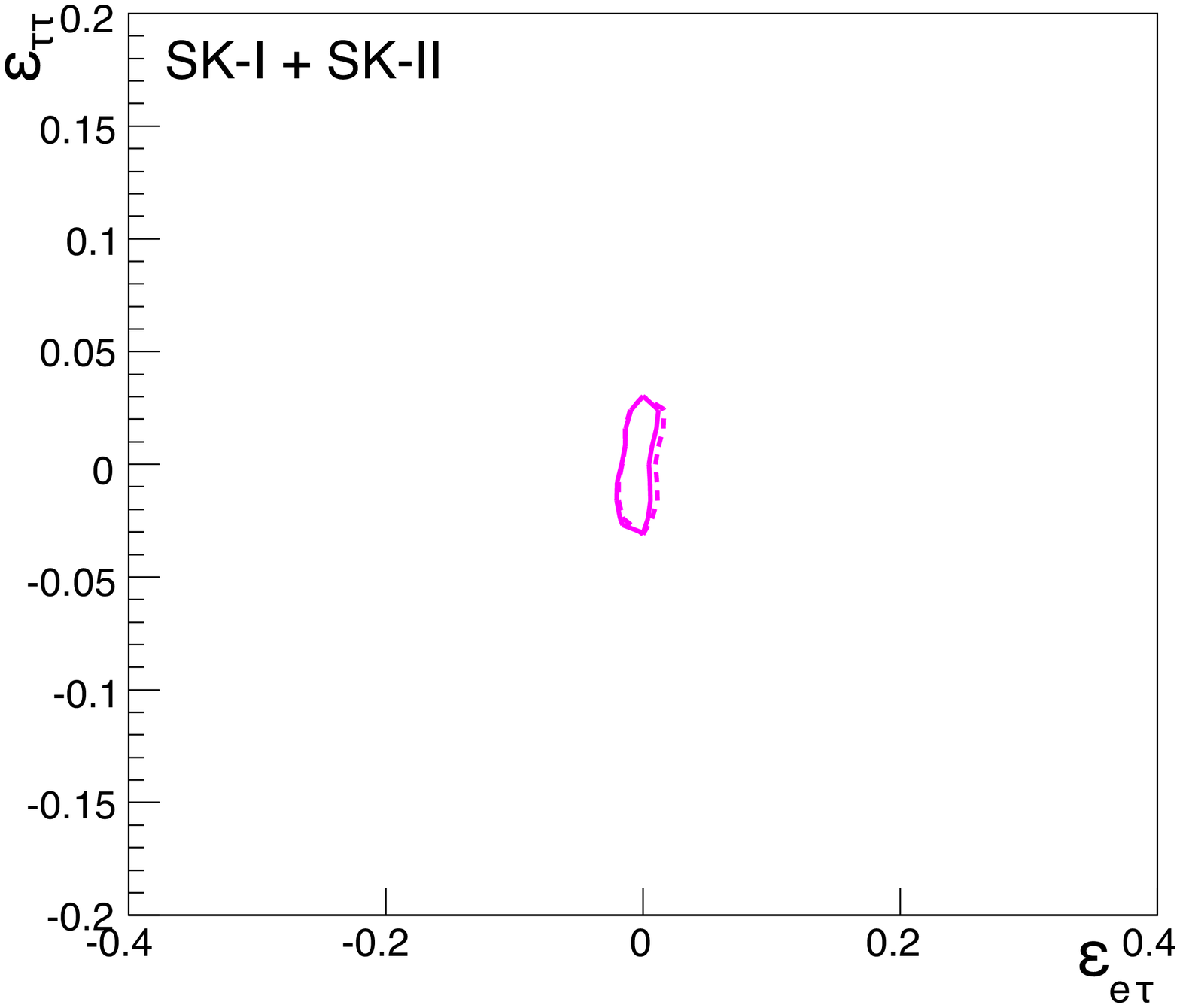,width=.30\textwidth,clip=}}
    \subfigure[$\eee=0.00$]{\epsfig{figure=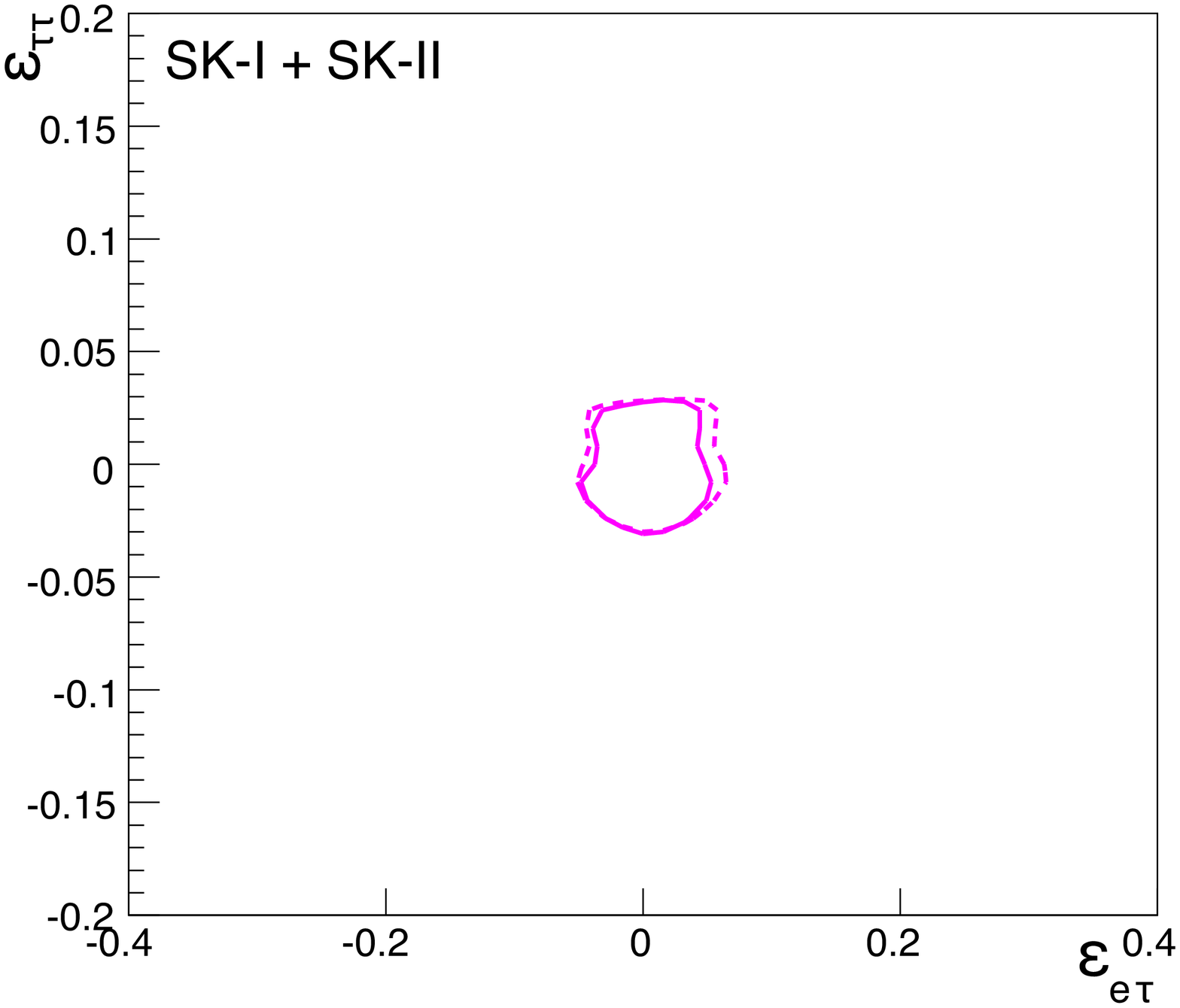,width=.30\textwidth,clip=}} \\
    \subfigure[$\eee=0.25$]{\epsfig{figure=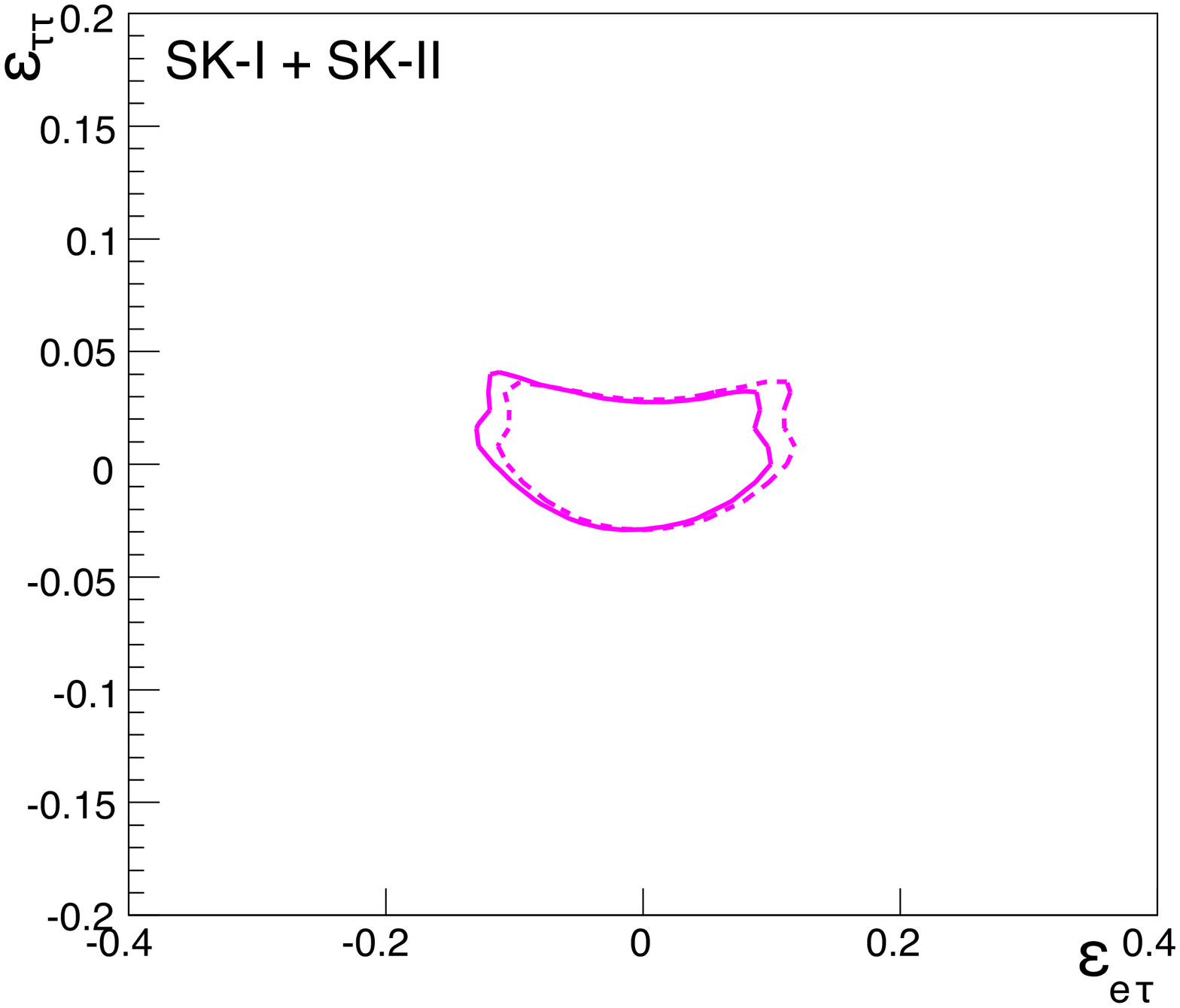,width=.30\textwidth,clip=}}
    \subfigure[$\eee=0.50$]{\epsfig{figure=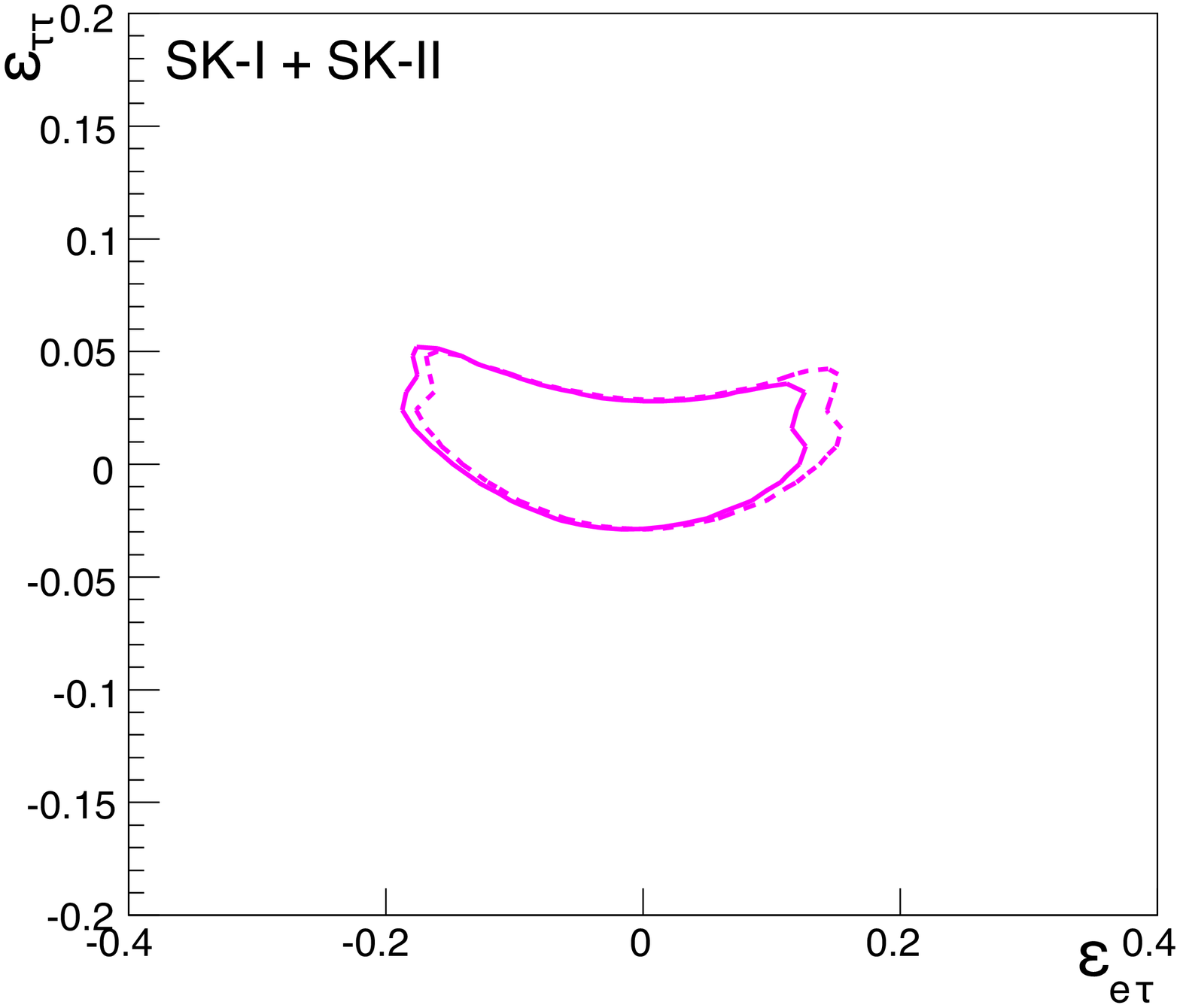,width=.30\textwidth,clip=}}
  \end{center}
    \caption{Allowed NSI parameters regions in the $\eet$ vs. $\ett$ plane at the 90$\%$ C.L. when  $\mbox{sin}^{2}\theta_{13} = 0.04$. 
             The solid curve and dashed curve indicate the normal inverted hierarchy fits, respectively.
            }
    \label{fig:3dnsi_allowed_theta13}
\end{figure*}

Present data suggest that $\theta_{13}$ is small relative to the 
other mixing angles, $\theta_{23}$ and $\theta_{12}$.
The Chooz experiment has placed the most stringent limit on the parameter, 
indicating that $\mbox{sin}^{2}\theta_{13} < 0.04 $~\cite{CHOOZ}.
Due to its small size its effects were ignored in the main analysis.
However, $\theta_{13}$ is expected to induce 
$\nu_{\mu} \rightarrow \nu_{e}$ transitions at multi-GeV energies 
possibly producing an 
excess of $\nu_{e}$ that could be misinterpreted as the effect of $\eet$. 
Here the analysis of section~\ref{chap:3d-hybrid} is repeated for 
the normal hierarchy, $\Delta m^2_{23} > 0$, and inverted hierarchy, $\Delta m^2_{23} < 0$, 
with $\theta_{13}$ fixed at the Chooz limit. 
Including $\theta_{13}$ breaks the positive-negative symmetry of $\eet$ so the 
fit is expanded to cover both regions. The results of the fit are again presented as a 
function of $\eee$ in Fig.~\ref{fig:3dnsi_allowed_theta13}.

Focusing on the normal hierarchy case, the addition of $\theta_{13}$ to the fit tends to improve the constraint 
on $\eet$.  The $\nu_{\mu} \rightarrow \nu_{e}$ oscillation probability in constant density 
matter is proportional to $\mbox{sin}^{2} \Theta_{13}$, where $\Theta_{13}$ is the effective $\theta_{13}$ mixing  
angle given by, 
\begin{equation}
  \Theta_{13} \sim \theta_{13} + \phi, \hspace{0.5cm} \tan2\phi \sim \frac{a\sin2\theta_{13}}{\Delta m^2_{31} -a\cos2\theta_{13}}.
  \label{eq:effective_theta13}
\end{equation}
\noindent In this equation $a$ is the product of the MSW matter potential and the neutrino energy, $\pm 2\sqrt{2}G_FN_eE_\nu$,
where the sign is positive (negative) for neutrinos (antineutrinos). 
The structure of the denominator can create a resonant enhancement of the oscillation probability depending upon 
the energy of the neutrino and the density of matter it traverses.
If $\theta_{13}$ is nonzero, the upward-going $\nu_{e}$ flux is expected 
to increase in the 2-10 GeV range, coinciding with the region where $\eet$ can enhance the $\nu_{e}$ flux. 
Accordingly, part of the $\nu_{e}$ flux induced by $\eet$ is now occupied by events from $\theta_{13}$ transitions
which therefore results in a tighter constraint on the parameter. 
However, the resonance behavior of Eqn.(\ref{eq:effective_theta13}) is contingent upon the 
signs of the mass hierarchy and MSW matter potential. 
This, coupled with unequal neutrino and antineutrino fluxes in the atmospheric data,
results in the asymmetric constraint on $\eet$ seen in Fig.~\ref{fig:3dnsi_allowed_theta13}.

\section{Conclusions}\label{chap:conclusions}

We have studied non-standard neutrino interactions in the context of atmospheric neutrinos
propagating in the Earth. Two analyses were presented considering possible effects from 
both flavor changing neutral current and lepton universality violating interactions.
Analysis of the SK-I and SK-II atmospheric neutrino data shows no evidence of NSI and provides the following constraints.
For NSI in the $\numu - \nutau$ sector, the 2-flavor hybrid model allows contributions from NSI in the form
\begin{align}
  &|\emt| < 1.1 \times 10^{-2}& \mbox{and} \nonumber \\  
  -4.9 \times 10^{-2} < &\ett - \emm < 4.9 \times 10^{-2}&,
\end{align}
\noindent at the 90\% C.L., where $\varepsilon$ and $\varepsilon'$ are replaced with $\emt$ and $\ett - \emm$, respectively.
In the 3-flavor hybrid model, the allowed regions are presented for different values of $\eee$ since the 
atmospheric data have no ability to constrain this parameter.


%
%
\section{Acknowledgements}\label{chap:acknowledgements}
We gratefully acknowledge the cooperation of the Kamioka Mining and Smelting Company. The SuperKamiokande experiment has been built and operated from funding by the Japanese Ministry of Education, Culture, Sports, Science and Technology, the United States Department of Energy, and the U.S. National Science Foundation. Some of us have been supported by funds from the Korean Research Foundation (BK21), and the Korea Science and Engineering Foundation. Some of us have been supported by the State Committee for Scientific Research in Poland (grant1757/B/H03/2008/35).

\end{document}